\documentclass{article}
\usepackage{arxiv}

\usepackage{rotating}

\usepackage{booktabs} 
\usepackage{lscape}
\usepackage{amssymb}
\usepackage{pifont}
\newcommand{\cmark}{\ding{51}}%
\newcommand{\xmark}{\ding{55}}%
\expandafter\let\csname equation*\endcsname\relax
\expandafter\let\csname endequation*\endcsname\relax
\usepackage{amsmath}
\usepackage{subcaption}
\usepackage{adjustbox}
\usepackage{threeparttable}
\usepackage[T1]{fontenc}
\usepackage{paralist}
\usepackage[breaklinks=true]{hyperref}
\usepackage{breakcites}

\title{A Survey on Bilateral Multi-Round Cloud-SLA Negotiation Strategies}

\author{Benedikt Pittl\\
	Faculty of Computer Science\\
	University of Vienna\\
	A-1090 Vienna, Austria\\
	\texttt{benedikt.pittl@univie.ac.at} \\
	\And
	Werner Mach\\
	Faculty of Computer Science\\
	University of Vienna\\
	A-1090 Vienna, Austria\\
	\texttt{werner.mach@univie.ac.at} \\
    \AND
    Erich Schikuta\\
	Faculty of Computer Science\\
	University of Vienna\\
	A-1090 Vienna, Austria\\
	\texttt{erich.schikuta@univie.ac.at} \\
}

\begin{document}

\date{}

\maketitle

\begin{abstract}
Today, static cloud markets where consumers purchase services directly from providers are dominating. Thus, consumers neither negotiate the price nor the characteristics of the service. In recent years, providers have adopted more dynamic trading mechanisms, as e.g. Amazon's EC2 platform shows: In addition to the reservation marketspace and the on-demand marketspace, Amazon offers a spot marketspace where consumers can bid for virtual machines. This spot marketspace was extended with spot blocks, and recently Amazon reworked the bidding options. In addition, other cloud providers, such as Virtustream, adopt dynamic trading mechanisms. The scientific community envisions autonomous multi-round negotiations for realizing future cloud marketspaces. Consequently, consumers and providers exchange offers and counteroffers to reach an agreement. This helps providers increase the utilization of their datacenters, while consumers can purchase highly customized cloud services.

In the paper at hand, we present a survey on multi-round bilateral negotiation strategies for trading cloud resources. Thus, we analyzed peer-reviewed articles in order to identify trends, gaps, similarities, and the scope of such negotiation strategies. In addition, we surveyed the formalism that the scientific community uses to describe such strategies. Based on these findings, we derived recommendations for creating and documenting bilateral multi-round negotiation strategies to foster their implementation in the industry.

\end{abstract}

\section{Introduction}
\label{sec:intro}

Today, cloud markets are emerging where cloud providers strive to develop and employ innovative vendor strategies to gain market shares~\cite{MessinaMMR17}. The leading platform for purchasing infrastructure as a service (IaaS) in the form of virtual machines (VMs) is Amazon Web Services (AWS), see e.g.~\cite{noauthor_companies_2017}. The traded virtual machines are preconfigured, called instance types on Amazon, and are sold on different EC2 marketspaces. Amazon distinguishes between
\begin{inparaenum}[(i)]
	\item a reservation marketspace,
	\item an on-demand marketspace and,
	\item a spot marketspace.
\end{inparaenum}
On the reservation marketspace, consumers have a long-term contract (e.g., three years) with Amazon. Such a long-term contract does not exist on the on-demand marketspace where virtual machines are charged per hour. 
In the spot market, consumers define the maximum price they are willing to pay per hour for a certain virtual machine. If the defined maximum price is higher than the so called~\emph{spot market price} then the consumer can use the virtual machine. If the spot market price exceeds the maximum price, then the virtual machine cannot be used. The spot market price is a dynamic price reflecting Amazon's current demand and supply. In 2017, Amazon reworked the spot marketspace: consumers need not bid anymore - an interesting comparison between the~\emph{old} spot marketspace and the~\emph{new} marketspace is given in~\cite{Chhetr18}. Amazon additionally offers so-called spot blocks. These virtual machines have a predefined lifetime. In addition to the three marketspaces, Amazon hosts a platform where consumers are able to sell virtual machines, which are purchased on the reservation marketspace, to other consumers.
While providers such as Amazon, Rackspace, and Microsoft Azure offer preconfigured virtual machines to consumers, other providers such as AT\&T cloud and Google allow consumers to purchase customized virtual machines~\cite{ChichinVK14}. The cloud provider Virtustream goes one step further~\cite{virtusteam}: It takes the notion of utility computing literally and charges consumers based on used $\mu$VMs, which are intended as a metric for measuring computational resources\footnote{see http://www.virtustream.com/software/micro-vms}. While a variety of different pricing models emerged for trading IaaS, Software as a Service (SaaS) providers such as Salesforce and Microsoft primarily charge consumers based on a monthly or annual fee.

Until now, neutral market platforms for trading cloud resources such as envisioned by~\cite{ChichinVK17,WALLOM19} failed in reality: The~\emph{Deutsche Boerse Cloud Exchange} was intended as such a platform which started in 2015 and closed in 2016. It helped consumers to compare the prices of VMs from different providers before purchasing them. In the scientific community, reasons for the failure of the platform, which was inter alia founded by the German Stock Exchange, have not been discussed yet. In industry-related literature, such as in~\cite{Herrmann2016}, the low technical maturity of the platform was mentioned as a reason. The limited number of cloud providers, which participated in the platform, is considered a further cause of the platform's failure: In total, three from Germany and one from the Czech Republic participated. 


The scientific community proposed over time different visions of future cloud markets, identifying economic aspects even in Grids~\cite{weishaupl2005business, schikuta2005business}, the computing paradigm preceding Clouds. 

Trust and security are critical enablers for the digital economy because they foster confidence among participants, ensuring that individuals and businesses are willing to engage in digital transactions and share sensitive data. Without trust~\cite{weishaupl2006gset}, users would hesitate to adopt digital platforms, and without security~\cite{shaaban2019ontology}, the risk of fraud, data breaches, and cyberattacks would undermine the integrity of the ecosystem. Digital auction mechanisms play a key role in guaranteeing trust and security by implementing transparent bidding processes, robust encryption, and secure authentication protocols. Transparency ensures fairness by allowing participants to see how bids are placed and winners are determined, reducing the risk of manipulation. Encryption protects sensitive data, such as payment information and bidding history, from unauthorized access, while secure authentication methods verify the identity of participants, preventing fraud. Together, these features create a trusted and secure environment that encourages participation and drives the growth of the digital economy.

Negotiation mechanisms, allowing for economic processes in the digital economy, range from centralized auctions~\cite{SamimiTM16} over decentralized auctions~\cite{bonacquisto_strategy_2014} to bilateral multi-round negotiations, which are often called bazaar-based negotiations~\cite{dastjerdi_autonomous_2015,PittlMS16}. The paper at hand focuses on autonomous bazaar-based negotiations where consumers and providers exchange offers until all offers are rejected or an agreement is formed. 
During the negotiations, consumers and providers use negotiation strategies.  The first negotiation strategy for negotiations on cloud markets, which considers prices and costs, was published in 2010~\cite{ShojaiemehrRQ18}. Indeed, autonomous negotiation is already well-established in the computer science domain. For instance, the Layer-2 protocol $PPP$ (Point-to-Point Protocol) foresees an autonomous negotiation of network configuration parameters. In contrast to negotiation approaches on cloud markets, these negotiations are not driven by prices and costs. In this paper, we focus on an analysis of recent negotiation strategies for cloud markets that are proposed by the scientific community.


The paper is structured as follows: Foundations of cloud markets as well as related work are introduced in section~\ref{sec:foundations}. The research questions and the applied methodology is introduced in section~\ref{sec:rq} followed by a summary of widely used negotiation protocols in section~\ref{sec:protocol}. The negotiation process is systematically evaluated along its three main steps: utility evaluation in section~\ref{sec:utility evaluation}, decision making in section~\ref{sec:decisionMaking}, and generation of counteroffers in section~\ref{sec:counteroffer}. The scope of the negotiation strategies introduced in scientific papers is presented in section~\ref{sec:scope}. A discussion of potential quality issues and biases of the survey is given in section~\ref{sec:quality}, followed by recommendations of the authors, which are presented in section~\ref{sec:discussion}. In section~\ref{sec:conclusion}, the findings of the survey are summarized.

\section{Foundations and Related Work}
\label{sec:foundations}
This section contains two parts. In the first part of this section, we introduce foundations, while the second part summarizes related work.

\begin{figure}
 \begin{center}
    \includegraphics[width=0.65\linewidth]{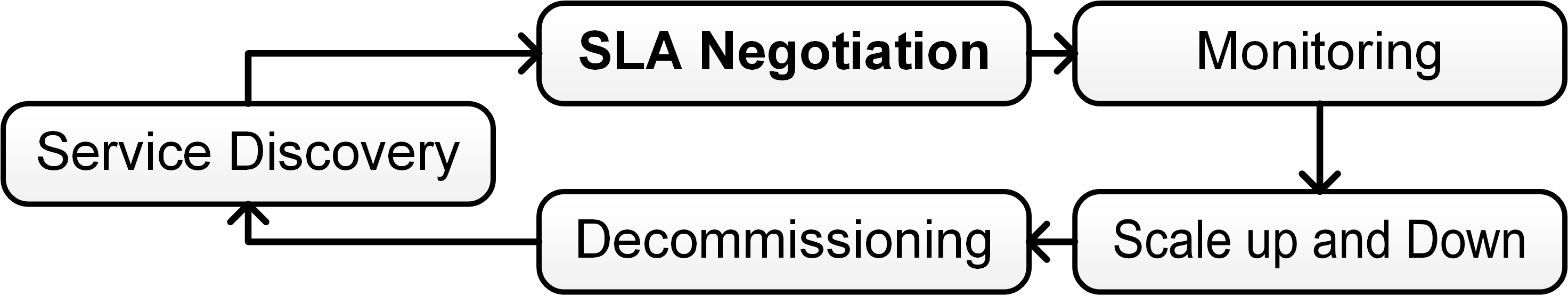}
\caption[SLA lifecycle]{SLA lifecycle based on~\cite{dastjerdi_autonomous_2015}}
     \label{fig:slaLifecycle}
 \end{center}
\end{figure}

Cloud services are usually described by SLAs, which specify the service characteristics by service level agreements~\cite{vinek2011classification}. SLA lifecycles such as illustrated in figure~\ref{fig:slaLifecycle} show the most important phases of SLAs from service discovery to decommission.
SLA negotiation is a separate step of SLA lifecycles~\cite{AshokM15,DastjerdiB15,DastjerdiB12,hasselmeyer2007implementing,HafezE16,joshi2009integrated}. This is because full autonomous bilateral multi-round SLA negotiation - in the rest of the paper termed~\emph{bilateral negotiation} - is a non-trivial task. Indeed, according to~\cite{ZhengMPB10} and~\cite{DastjerdiB12}, SLA negotiation is the most complex task of the SLA lifecycle. The authors of~\cite{DastjerdiB12,GhummanS17} consider the negotiation of SLAs as the bottleneck of the SLA lifecycle because this step still requires human intervention, making the negotiation time-consuming. Additionally, human intervention in negotiation is error-prone due to irrational actions of them~\cite{GhummanSL16,alsrheed2014intelligent}. This leads to SLAs that do not meet the business needs of consumers - consequently, autonomous negotiation mechanisms are required~\cite{HollowaySS15}.  Autonomous negotiations can reduce SLA violations and negotiation failures~\cite{SonKHKC16}.
Especially if consumers have special requirements on a service, then non-negotiated SLAs might not be appropriate, and so autonomous bilateral negotiations can significantly improve the utility of SLAs~\cite{rajavel2015optimizing}.
Also, the high volatilities on the cloud market, which are higher than those of stock exchanges, emphasize the need for autonomous bilateral negotiation mechanisms for trading cloud resources~\cite{IshikawaF15}.  Dynamically varying objectives and requirements are further characteristics of Cloud services, which makes their autonomous negotiation on cloud markets necessary~\cite{AshokM15}.
In this paper, we follow the definition of~\cite{ItoHK07} who describe autonomous negotiation as~\emph{a technology to
automatically search an agreeable point on the given utility spaces of each
(software) agent, without unnecessarily revealing the utility spaces as little as possible
based on a negotiation protocol, and often applied to multi-issue agreement
problems.}

\begin{figure}
    \centering
    \begin{subfigure}[b]{0.59\textwidth}
        \centering
        \includegraphics[width=1\linewidth]{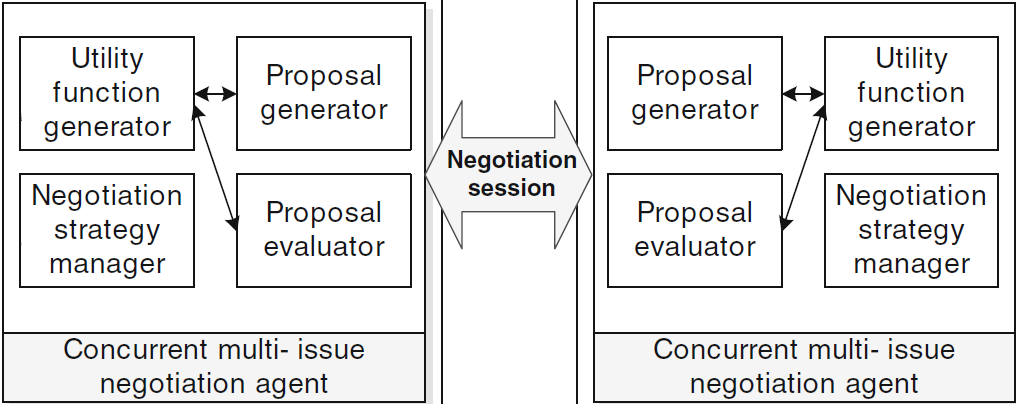}
        \caption{Excerpt of framework introduced in~\cite{SonS15}}
        \label{fig:framework1}
    \end{subfigure}
    \hfill
        \begin{subfigure}[b]{0.39\textwidth}
        \centering
        \includegraphics[width=0.85\linewidth]{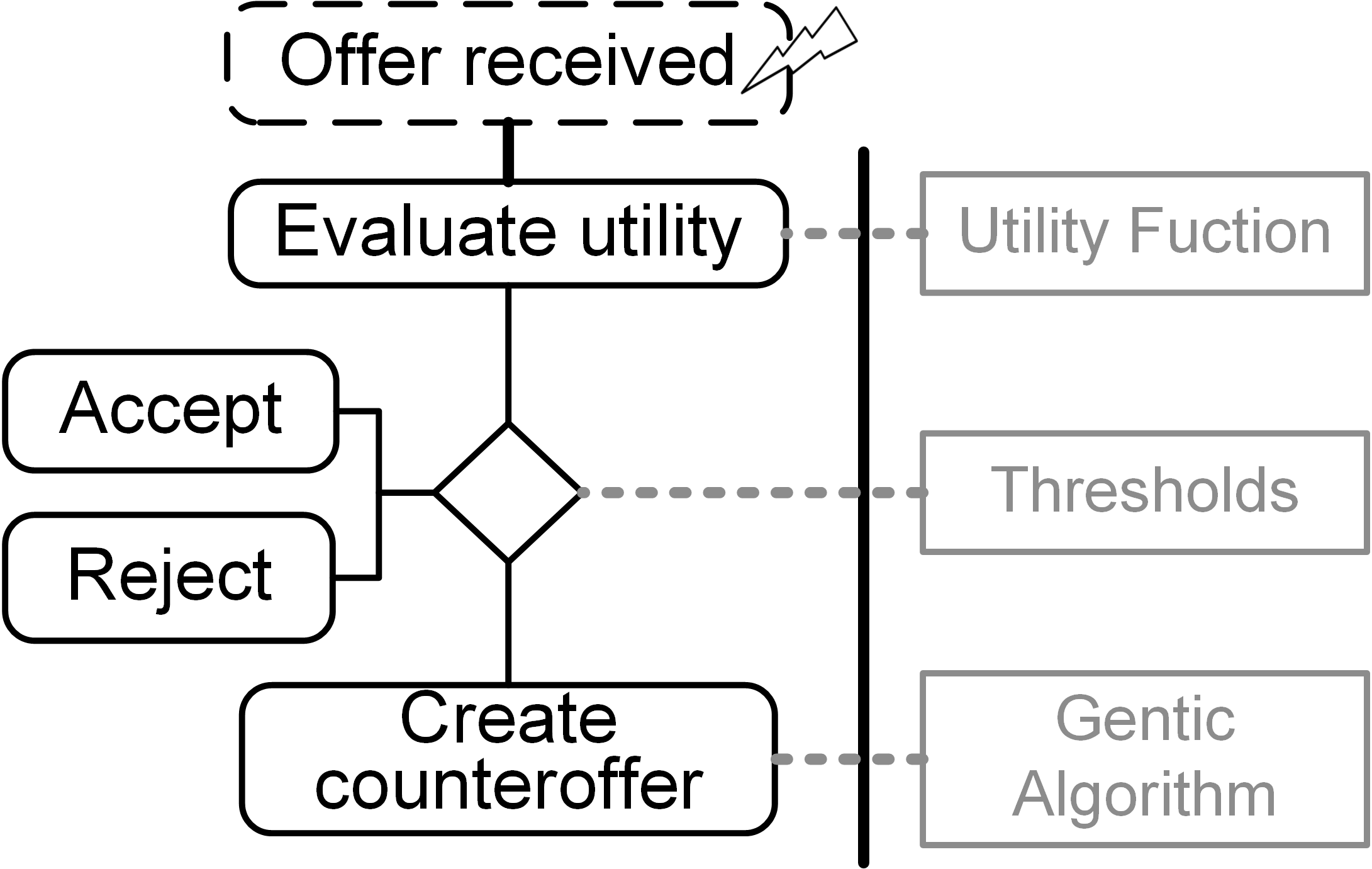}
        \caption{Excerpt of framework introduced in~\cite{PittlMS16}}
        \label{fig:framework2}
    \end{subfigure}
    \begin{subfigure}[b]{0.55\textwidth}
        \centering
        \includegraphics[width=1\linewidth]{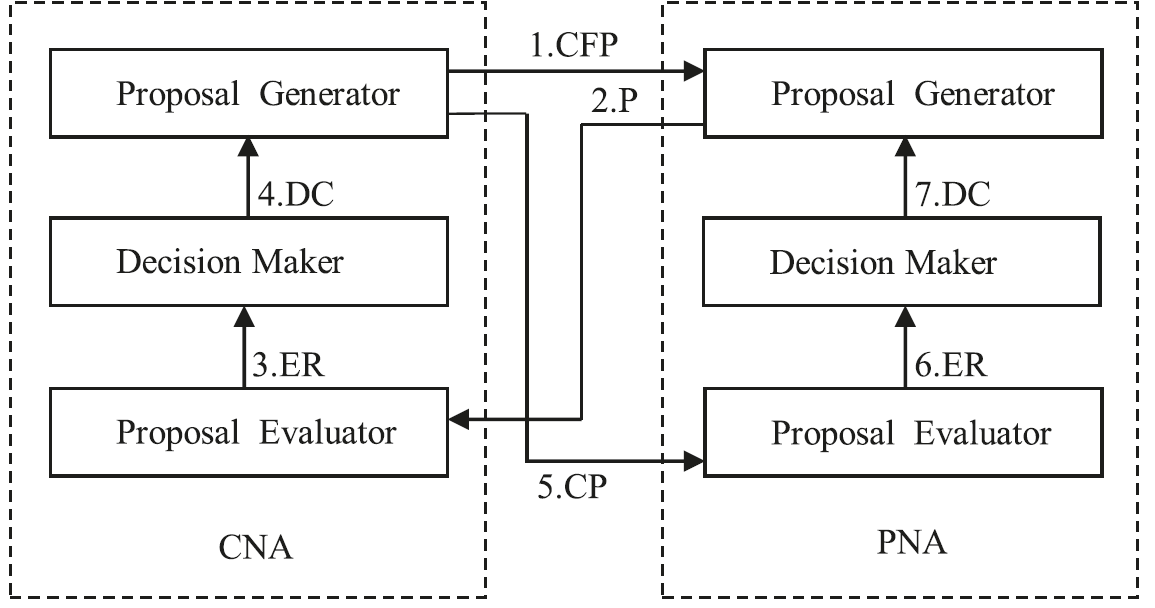}
        \caption{Framework introduced in~\cite{ShojaiemehrRQ18}}
        \label{fig:framework3}
    \end{subfigure}
    \hfill
    \begin{subfigure}[b]{0.44\textwidth}
        \centering
        \includegraphics[width=0.95\linewidth]{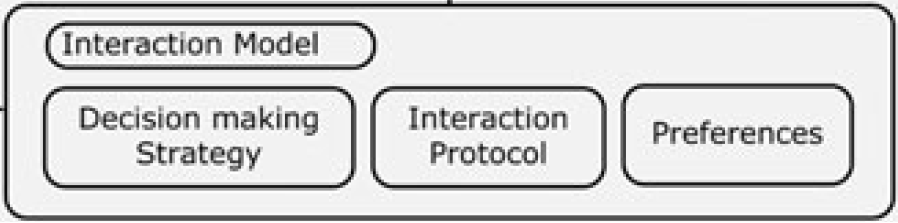}
        \caption{Excerpt of framework introduced in~\cite{baruwal2015autoslam}}
        \label{fig:framework4}
    \end{subfigure}
    \caption{Excerpt of negotiation frameworks}
    \label{fig:negotiationFramework}
\end{figure}

The negotiation process can be structured along three steps~\cite{ShojaiemehrRQ18,PittlMS16,PittlMS162}:
\begin{inparaenum}[(i)]
	\item Evaluation,
	\item Decision, and
	\item Counteroffer generation.
\end{inparaenum}
In the evaluation step, the received offers are evaluated and ranked. In the decision step, the strategy decides how to respond to offers, for example, with an agreement, while in the last step, counteroffers are generated - if the negotiation was not terminated in the decision step. Figure~\ref{fig:negotiationFramework}  shows four exemplary negotiation frameworks of scientific publications - all of them encompass components which support these three steps. The framework depicted in figure~\ref{fig:framework1} was introduced by~\cite{SonS15} and explicitly emphasizes the~\emph{Utility Function Generator} component. It is used for the counteroffer generation (Proposal generator in figure~\ref{fig:framework1}) and for the evaluation (Proposal evaluator in figure~\ref{fig:framework1}). Additionally, the framework foresees a negotiation strategy manager who is responsible for choosing the appropriate negotiation strategy as well as the setup. Figure~\ref{fig:framework2} shows a framework that foresees the three aforementioned steps. Also, here utility functions are suggested to do the evaluation of the offers. Almost the same framework was introduced by Shojaiemehr et al.~\cite{ShojaiemehrRQ18} - which is depicted in figure~\ref{fig:framework3} - and by authors of~\cite{CoutinhoCSGJ16} who see negotiation as a set of complex actions such as the creation of a new proposal, evaluation of them and making a decision about their acceptance or rejection. In the framework, which figure~\ref{fig:framework4} shows, the preferences represent the utility functions. Also, a decision-making process is part of the framework. Missing is the counteroffer generation approach. Instead, the negotiation protocol (interaction protocol) - which is discussed in a separate part of the paper at hand - is explicitly mentioned. Also, Ghumman et al. see the negotiation protocol as part of the negotiation strategy by mentioning the following four components of their negotiation framework:
\begin{inparaenum}[(i)]
	\item a negotiation protocol
	\item a negotiation strategy
	\item a concession computation function for creating counteroffers as well as a
	\item an offer generation method using utility functions~\cite{GhummanSL16}.
\end{inparaenum}
Also in related work, we were able to identify the aforementioned three negotiation steps, even if the terms used are different. For instance, Baruwal et al. describe that consumers and providers~\emph{have to make decisions about what initial offer to make, what counter-offer to make, whether to accept an offer, and when to terminate negotiation}~\cite{baruwal2015autoslam}. Similarly, Yaqub et al. describe that
\begin{inparaenum}[(i)]
	\item a bidding function that generates counteroffers
	\item an opponent model for modeling the opponent and
	\item a decision maker which decides when to make a bid~\cite{YaqubYWKLJ14} is part of a negotiation strategy.
\end{inparaenum}
Due to the dominance of the aforementioned three negotiation steps, we organized our paper along them. Additionally, we surveyed the underlying negotiation protocols.


While autonomous negotiation is a long-pursued vision in Computer Science~\cite{ShojaiemehrRQ18}, the scientific community mentions specific characteristics of them in the context of cloud markets.
The dynamic resource utilization of datacenters for the pricing of offers is considered as the main influence factor of bilateral negotiations on cloud markets~\cite{BaranwalKRV18}. This is in line with Dastjerdi et al.~\cite{DastjerdiB15} who see the utilization of datacenters as a~\emph{key aspect} for autonomous negotiations in the cloud domain. Also, the authors of~\cite{SonS15} see overprovisioning and underprovisioning of cloud providers as a significant motivation to apply dynamic negotiation approaches.  In addition to the dynamic workload, Rajavel et al. and Son et al. see the characteristics of cloud services as an aspect that has to be considered during autonomous bilateral negotiations on cloud markets~\cite{RajavelT16,SonKHKC16}.  Latter summarizes the under- and over-provisioning issue under the term~\emph{dynamic workflow}.
The cloud services that providers offer are different according to server space, locations, availability, and security~\cite{IshikawaF15}.
Baranwal et al. mention the non-cooperation and self-interest as two main pillars for negotiations in cloud markets~\cite{BaranwalKRV18} - without detailing how these two pillars are specific for cloud markets.

Publicly available information of the negotiation partners is unrealistic but often assumed by approaches focusing on an optimal resource allocation using game theory~\cite {HollowaySS15}. 
Approaches which pursue the usage of multi-round bilateral negotiations - instead of game theoretical approaches - such as described in~\cite{HafezE16,BaarslagHHJ16,IlanyG16,PittlMS16,PittlMS16EDOC} consider a cloud market as a multi-agent market whereby each agent has incomplete information.
The lack of information negotiation is a well-known fact and will not be changed in the following years~\cite{HollowaySS15}. So the survey focuses on approaches that do not use game theory.

The scientific community published multiple papers about bilateral negotiations on cloud markets, which motivated us to conduct the analysis introduced in the paper at hand.
The authors of~\cite{ShojaiemehrRQ18} analyzed 25 papers about cloud service negotiation, which were published between 2010 and 2016. The survey considered not only bilateral cloud SLA negotiation strategies but also multilateral cloud SLA negotiation strategies. Further, single-round cloud SLA negotiation strategies, such as auctions, were considered. The authors classified the identified negotiation strategies along the cloud service types SaaS, PaaS, and IaaS. A focus on state-of-the-art negotiation strategies, as well as a deep analysis of the negotiation strategies, is missing in the paper. 

The paper~\cite{PittlMS16} introduces a classification of bilateral cloud SLA negotiation strategies. The authors present a generic cloud market ecosystem and suggest using autonomous bilateral multi-round negotiation strategies on such markets. The paper classified negotiation strategies along their scope. The authors concluded that existing negotiation strategies are incomplete and need to consider economic concepts in order to be adopted in industry.  A detailed analysis of the content of the negotiation strategies is missing.

Hani et al. introduced in~\cite{HaniPH15} a prominent survey on SLA management techniques. Thereby, the authors introduce SLA Management tools. SLA Renegotiation, SLA Violation Detection, and Prediction, as well as Penalty, are some of the salient points of cloud service management that the authors identified. The paper has a strong focus on SLA re-negotiation and therefore the authors classified existing approaches along the following categories: game theory, reinforcement learning, Bayesian learning, and rule-based approaches. A focus on bilateral multi-round negotiation strategies, as well as an analysis of them, is missing in the paper. However, the paper mentions counteroffer generation approaches for re-negotiations.

In~\cite{HafezE16}, a survey of different negotiation approaches is presented. The survey neither focuses on multi-round bilateral negotiations nor on cloud-SLAs. For instance, approaches based on auctions were surveyed, as well as non-cloud-based approaches. Hence, a specific focus on autonomous bilateral multi-round negotiation is missing.

The survey on opponent learning techniques presented in~\cite{BaarslagHHJ16} introduces generic approaches, neither focusing on the cloud domain nor on multi-round bilateral negotiations. Similarly, in~\cite{SinghC16} a generic review of cloud resource provisioning approaches is given without surveying existing scientific approaches. The survey presented in~\cite{wazir2016service} summarizes SLA architectures in the cloud computing domain without systematically surveying SLA negotiation strategies. Although some of the frameworks have a negotiation component, they are not systematically analyzed in the paper.

An overview of utility computing and cloud computing approaches is presented by~\cite{WuGB15}. The paper describes the SLA lifecycle and SLA management technologies. While SLA negotiation is mentioned, a detailed description of appropriate strategies is missing.

Manvi et al. published a survey on resource management techniques in clouds~\cite{ManviS14}. Thereby, the definition of resource management is broadly so that it inter alia includes allocation, provisioning, discovery, and estimation. Due to the broad scope of the survey, it does not encompass bilateral negotiation. Rather, it focuses inter alia on surveying performance metrics and resource-mapping schemas.

The authors of~\cite{ToosiCB14} present a generic survey on interconnected cloud computing environments. SLA negotiation is mentioned as an important step, but details or classifications of them are not presented in the paper. The survey presented in~\cite{AlhamadDC11} focuses on SLA performance measures in cloud computing but excludes SLA negotiation techniques.

The related work analysis shows that the scientific community surveyed a couple of SLA management techniques and frameworks. A first analysis of negotiation strategies is presented in~\cite{PittlMS16} and~\cite{HaniPH15}. However, a systematic classification and analysis of autonomous bilateral negotiations strategies, which analyzes, e.g., how offers are evaluated, is missing.

\section{Research Questions and Methodology}
\label{sec:rq}

For the definition of the research questions, we started an informal discussion with research peers to identify a relevant scope and focus of the survey. During the discussions, we extended and adapted research questions. Finally, we defined the following research questions and added the sections of the paper at hand that address these questions:
\begin{itemize}
	\item \textbf{RQ1:} What bilateral multi-round negotiation strategies are existing for cloud-SLAs?
	\begin{itemize}
		\item How are offers evaluated? $\mapsto$ Section: \ref{sec:utility evaluation}
		\item How is decision-making done? $\mapsto$ Section: ~\ref{sec:decisionMaking}
		\item How are counteroffers generated? $\mapsto$ Section: ~\ref{sec:counteroffer}
		\item What is the scope of these strategies? $\mapsto$ Section: \ref{sec:scope}
		\item Which algorithms or computational techniques are they using? $\mapsto$ Sections: \ref{sec:utility evaluation},~\ref{sec:decisionMaking} and~\ref{sec:counteroffer}	
		\item How are these negotiation strategies formalized? $\mapsto$ Section: \ref{sec:scope}
		\item What protocols are used for the negotiation strategies? $\mapsto$ Sections: \ref{sec:protocol}
	\end{itemize}
	\item \textbf{RQ2:} Are the negotiation strategies for bilateral multi-round negotiations implemented?
		\begin{itemize}
		\item Which frameworks/libraries are used for their implementation? $\mapsto$ Sections: \ref{sec:scope}
		\item How are they tested? $\mapsto$ Sections: \ref{sec:scope}
		\end{itemize}
\end{itemize}

Each of these two questions has a different focus. Research question 1 aims at identifying the scope of bilateral multi-round negotiation strategies as well as the algorithms used. Research question 2 aims at identifying eligible libraries and frameworks for their implementation. 

\begin{figure}[htbp]
 \begin{center}
    \includegraphics[width=0.99\linewidth]{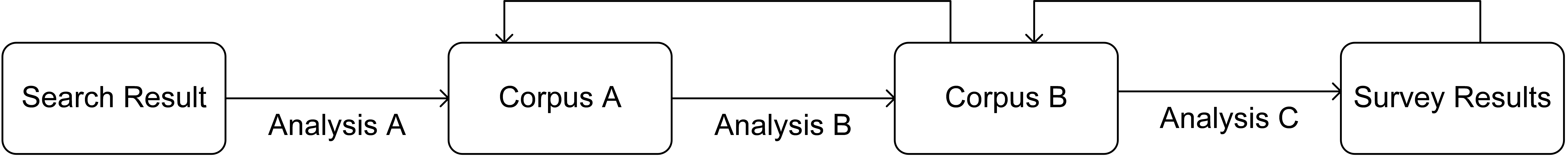}
    \caption{Survey process and creation of the corpses}
    \label{fig:surveyprocess}
 \end{center}
\end{figure}

The applied survey process is depicted in figure~\ref{fig:surveyprocess} and encompasses three analysis steps and three corpses. Table~\ref{tab:surveyCriteria}\footnote{the table is discussed in more detail in a following section} summarizes the process: Here we list the review methods, the exclusion/inclusion criteria, as well as the executors of the steps.  For the generation of the corpus~\emph{Search Result} we executed a keyword search using Google Scholar\footnote{scholar.google.com}, ACM Library\footnote{https://dl.acm.org/}, Science Direct\footnote{https://www.sciencedirect.com/} and IEEEXplore\footnote{https://ieeexplore.ieee.org/Xplore/home.jsp} during 02.06.2018 and 04.06.2018. The aim of the keyword search was to identify peer-reviewed scientific conference papers and journals in the domain of bilateral multi-round negotiations. After a discussion, we agreed to use the keywords~\emph{bilateral negotiation strategy},~\emph{cloud negotiation strategy}, and~\emph{bazaar negotiation strategy}. The first keyword is more generic than the second one, so that we do not a priori exclude papers about bilateral multi-round negotiations which do not emphasize their cloud-SLA specific focus. In the scientific community, the term~\emph{Bazaar} is often used as a synonym for bilateral multi-round negotiations. Hence, we decided to use the third keyword additionally. For the first two keywords, we considered  200 peer-reviewed scientific publications (100 for each keyword), while for the third keyword, 30 peer-reviewed scientific publications were considered. The highest ranked publications according to the relevance score\footnote{order of search result} were used to create the initial corpus~\emph{Search Result}.  In the analysis step $A$, only exclusion criteria were defined (see table~\ref{tab:surveyCriteria}) to create corpus $A$. To prevent depreciated results of the survey, we ignored publications that were published before 2014\footnote{We defined this constraint in search settings so that the corpus~\emph{Search Results} was not reduced by this constraint}. For the same reason, we ignored books as they usually require considerable time until publication. Additionally, patents were excluded. We also excluded papers that were obviously not relevant for the survey by reading the title, keywords, conference name/journal name, and by cross-reading the abstract. E.g., the publication~\cite{collyer2016geopolitics} was part of the search result but is definitely not relevant for the survey as the title of the paper reveals:~\emph{Geopolitics as a migration governance strategy: European Union bilateral relations with Southern Mediterranean countries}.

The so-generated Corpus A was refined during Analysis B: Each paper was reviewed again in order to identify its relevance for this survey. In our context, relevant means that the paper focuses on the description of a multi-round bilateral negotiation strategy for cloud-SLAs. Therefore, we read the abstract, the title, the title of the conference/journal, and the keywords. Additionally, we cross-read the content of the paper. Afterward, we classified the publications using three categories: I - relevant for this survey, II - not relevant for this survey, III - borderline, needs further discussion. The review, as well as the classification, was executed by two persons (one PhD student and one PostDoc) in parallel. Papers that belong to class III were reviewed in more detail.
A final classification was done after a discussion between all three authors. So, finally, each publication was assigned to one of the first two classes. Two exemplary papers that were initially assigned to class III are the following:
\begin{itemize}
\item
The work introduced in~\cite{chen2015cpn} was motivated by the assumption that due to the mass of credentials, the negotiation efficiency is not high. Hence, the authors developed a so-called~\emph {Automated Trust Negotiation} which manages digital certificates gradually between cloud providers and consumers.
\item
Macias et al. \cite{MaciasG16} introduced a decentralized reputation model. Market participants should be able to assess the reputation of other market participants, which influences bilateral multi-round negotiations.
\end{itemize}
As neither papers introduce a negotiation strategy, we finally assigned them to class II.
 This analysis process was executed iteratively, which is shown by the backloop in figure~\ref{fig:surveyprocess}. Finally, 26 peer-reviewed papers were identified as relevant for this survey (class I) and therefore used for answering the research questions. We read these papers and discussed them in order to gain insights into the research domain: Within Analysis C we executed a detailed analysis which is presented in the following sections of the paper at hand. The survey was executed by three persons (one PhD student, one PostDoc one Professor). 

\section{Protocols}
\label{sec:protocol}

The authors of~\cite{PittlMS16} and~\cite{ShojaiemehrRQ18} see the negotiation protocol as the main pillar for realizing autonomous negotiations. Shojaiemehr et al. define the negotiation protocol as a document which~\emph{specifies the structure and the context of messages circulated among the agents}~\cite{ShojaiemehrRQ18}. This is in line with the scope of current existing specification documents such as the WS-Agreement Negotiation specification~\cite{waeldrich_ws-agreement_2011}.  Baarslag et al. see the protocol as part of a so called~\emph{negotiation setting} which further encompasses the participants as well as the~\emph{negotiation scenario}~\cite{BaarslagHHJ16}. A negotiation scenario is a use case where each participant has a concrete negotiation strategy as well as a goal. Son et. al. summarized the negotiation protocol as a set of rules between negotiation parties~\cite{SonKHKC16}. Similarly,~\cite{baruwal2015autoslam} describe that negotiation protocols govern the interactions with a set of rules for the conversation.

\begin{table}
\caption{Summary of protocols}
\centering
\begin{tabular}{lp{7.8cm}}

\hline
\hline
& Papers using this Protocol \\
\hline
\hline
Rubinstein's Protocol & \cite{SonKHKC16,SonS15,alsrheed2014intelligent,YaqubYWKLJ14,NajjarBP17}  \\
WS-Agreement Negotiation  & \cite{PittlMS15,PittlMS16,PittlHMS17wetice} \\
Own Protocol & \cite{AbedinCG14,BaranwalKRV18,HollowaySS15,GhummanSL16,IshikawaF15,RajavelT16,rajavel2015optimizing,DastjerdiB15,CoutinhoCSGJ16,chen2014negotiation,
dhanasekaran2015dynamic,RajavelT162,baruwal2015autoslam,GhummanS17,AshokM15,GalI15,IlanyG16,rajavel2017adslanf}\\
\hline
\end{tabular}

\vspace{2em}
\label{tab:protocol}
\end{table}

\begin{figure}
    \centering
    \begin{subfigure}[b]{0.45\textwidth}
        \centering
    	\includegraphics[width=0.99\linewidth]{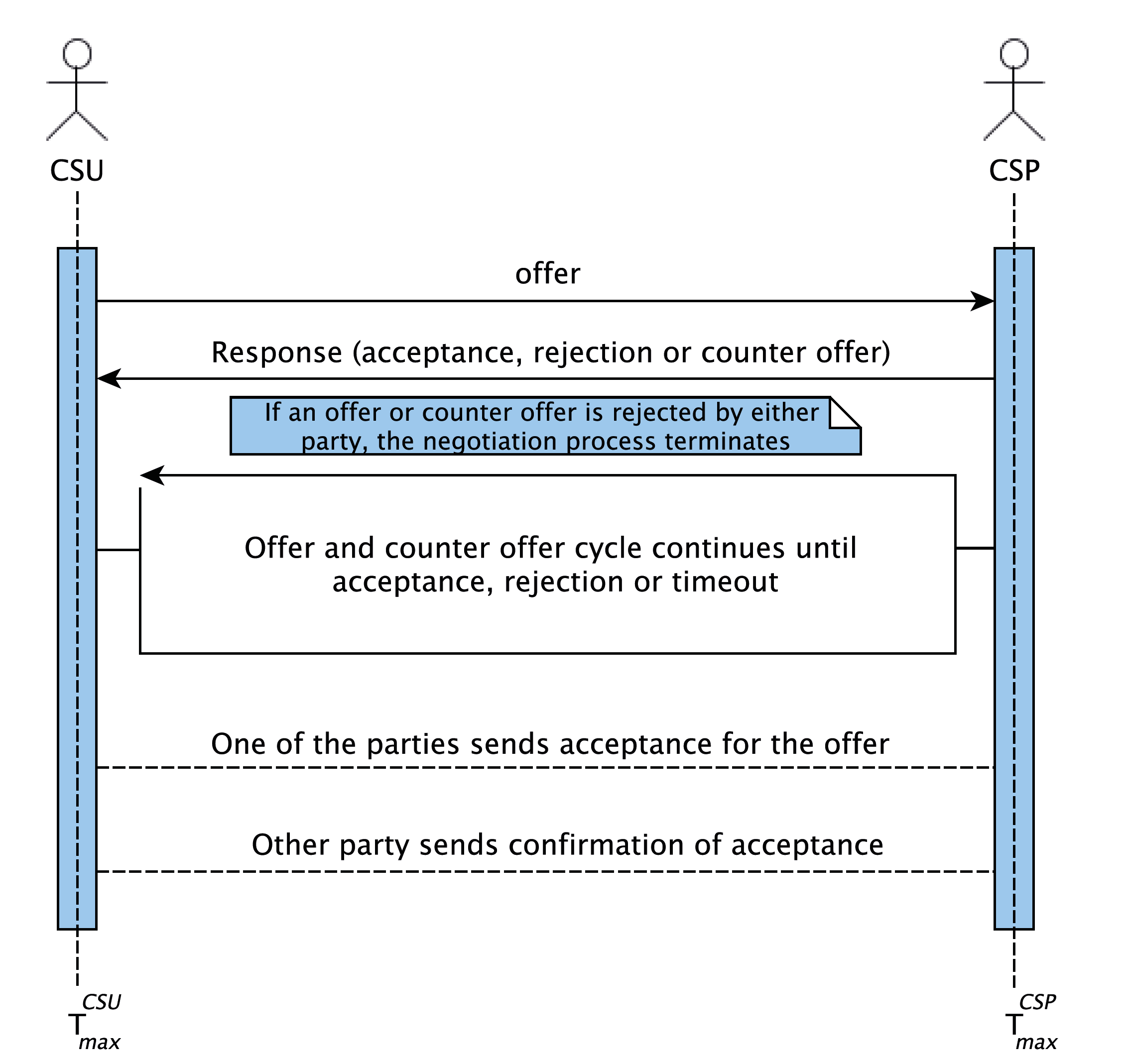}
    	\caption{Flip-Flop negotiation protocol introduced in\protect\cite{GhummanSL16}}
    	\label{fig:flipFlopProtocol}
    \end{subfigure}
    \hfill
    \begin{subfigure}[b]{0.45\textwidth}
        \centering
        \includegraphics[width=1\linewidth]{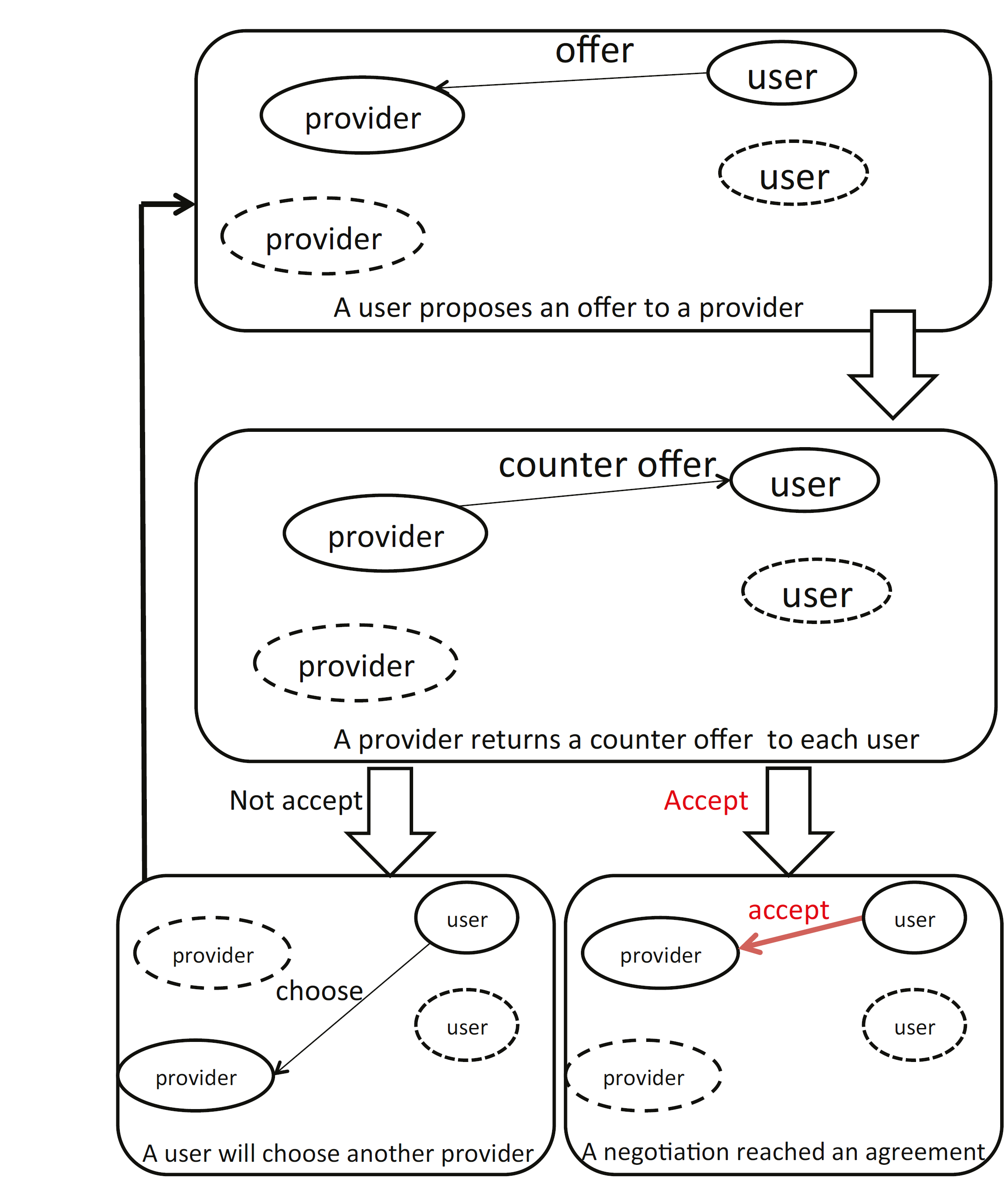}
        \caption{Negotiation protocol introduced in\protect\cite{IshikawaF15}}
        \label{fig:federatedProtocol}
    \end{subfigure}
    \caption{Excerpt of negotiation protocols}
    \label{fig:negotiationProtocols}
\end{figure}

Table~\ref{tab:protocol} summarizes the negotiation protocols used to describe the negotiation strategies. It shows that two protocols are dominating in the scientific community: Rubinstein's Protocol and the WS-Agreement Negotiation protocol. Rubinstein's alternating offer protocol was introduced in~\cite{rubinstein1982perfect}. In this work, Rubinstein described the protocol with an exemplary negotiation about how to divide a pie: \emph{Each has to make, in turn, a proposal as to how it should be divided. After one party has made such an offer, the other must decide either to accept it or to reject it and continue with the bargaining.} The widely used protocol is a bilateral protocol where two negotiation partners exchange alternately an offer. The receiver can accept the offer, reject it, or make a counteroffer. In the first case, an agreement is formed, while in the second case, the receiver of the counteroffer can accept the received offer or make a counteroffer. An extended Rubinstein protocol in the cloud-SLA context is depicted in the sequence diagram in figure~\ref{fig:flipFlopProtocol}. $CSU$ stands for cloud service user while $CSP$ stands for cloud service provider. Both have a negotiation deadline abbreviated with $T_{max}$. Here, an additional confirmation of an acceptance message is foreseen - in Rubinstein's protocol, it is missing. Indeed, most of the papers do not argue why Rubinstein's protocol is used. Only~\cite{alsrheed2014intelligent} describes that it was used because of its simplicity. Son et al. describe that it was used because it is generic and emulates real-life negotiation~\cite{SonS15}.

\begin{figure}
 \begin{center}
    \includegraphics[width=0.5\linewidth]{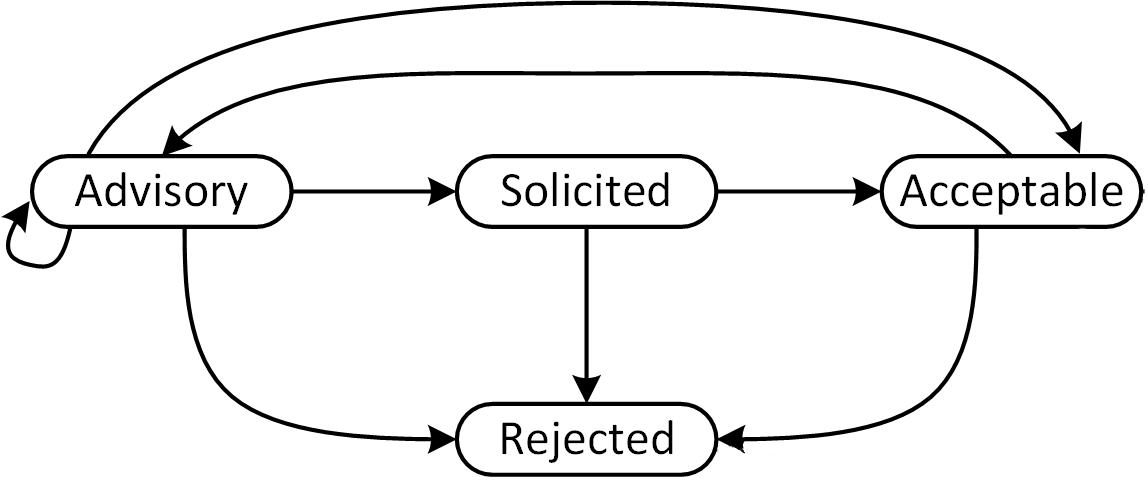}
    \caption{Transition diagram of the WS-Agreement Negotiation Specification\protect\cite{waeldrich_ws-agreement_2011}}
    \label{fig:ws_agreement_offers}
 \end{center}
\end{figure}

Papers such as~\cite{PittlMS15,PittlMS16,PittlHMS17wetice} suggest using the protocol described in the WS-Agreement Negotiation specification.
The WS\footnote{WS stands for Web Service}-Agreement Negotiation specification~\cite{waeldrich_ws-agreement_2011} is managed by the Open Grid Forum and aims at specifying negotiations for web services. It is an extension of the WS-Agreement specification~\cite{andrieux_web_2007} and describes an XML-based structure of offers as well as their possible states. In total, the WS-Agreement Negotiation specification defines four states of offers. These four states and their transitions are illustrated in figure~\ref{fig:ws_agreement_offers}. An offer in the advisory state requires further negotiation as it is, e.g., not completely specified. The solicited state is used for offers that are completely specified. The negotiation party that receives such an offer is forced either to accept the offer so that the state of the offer becomes acceptable or reject it, which leads to the state being rejected. Acceptable offers might result in agreements. Agreements are offers to which consumers and providers agree. As described in the following section, offers in the acceptable state of the WS-Agreement Negotiation standard are not binding. ~\emph{The  ACCEPTABLE  state indicates that a negotiation participant is willing to accept a negotiation offer as is.} But it is also described that~\emph{there is no guarantee that a subsequent agreement is
created}.  Hence, works such as~\cite{Mach17} introduced further states.
 The rejected state is used for offers that are rejected.

\begin{figure}
 \begin{center}
    \includegraphics[width=0.65\linewidth]{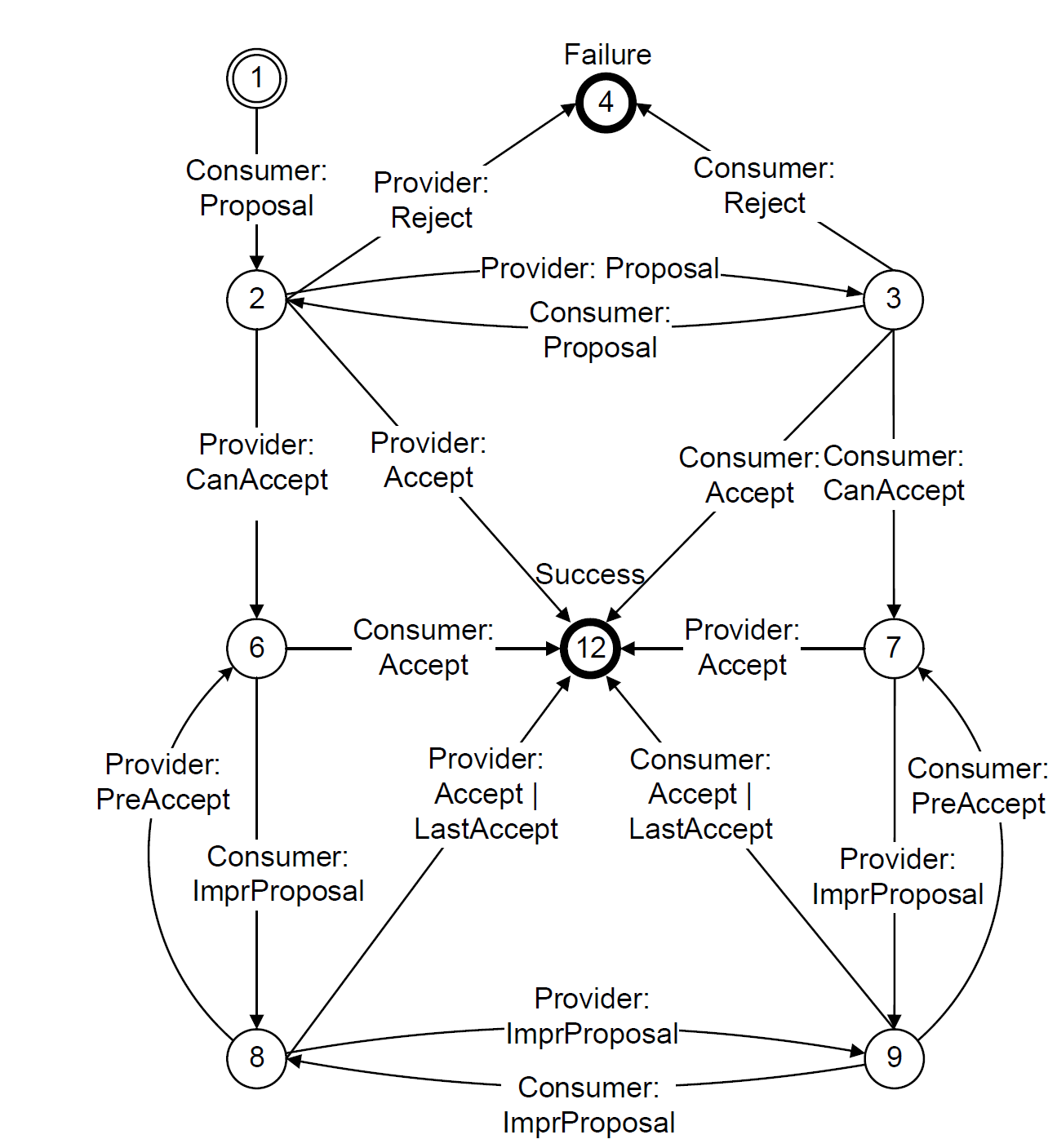}
    \caption{State diagram of negotiation protocol introduced in\protect\cite{HollowaySS15}}
     \label{fig:HolloProtocol}
 \end{center}
\end{figure}

Most of the papers introduced their own protocol.
So the authors of~\cite{HollowaySS15} introduce a novel protocol for their negotiation strategy, which is depicted in figure~\ref{fig:HolloProtocol}. The state diagram shows that 7 different types of messages exist: Proposal, Reject, CanAccept, ImprProposal, PreAccept, LastAccept, and Accept. For more details see~\cite{HollowaySS15}. Not all protocols that we assigned to this category are completely novel. Indeed, they are based on existing specifications. So, e.g.~\cite{GhummanSL16} introduced their own protocol where providers offer so-called templates which contain parameters used as a starting point for the negotiation process. The authors state that this protocol is an extension of Rubinstein's negotiation protocol. A further difference is that the protocol limits the negotiation time instead of the number of negotiation rounds. This has the benefit that the deadline of the negotiation process can be described more precisely. Also, the protocol used in~\cite{IshikawaF15} is based on Rubinstein's protocol, even if not mentioned explicitly. It is described that the negotiation partners exchange offers. If the negotiation partner uses the received offer as a counteroffer, an agreement is formed. Otherwise, it can look for a provider from the list and restart the complete process. So the negotiation process itself encompasses at most three exchanges of offers. Negotiation stops if the deadline is reached - a reject message is not described. An example is illustrated in figure~\ref{fig:federatedProtocol}: First, the user sends an offer to a provider, which responds with a counteroffer. If the counteroffer is identical to the offer, an agreement is formed; otherwise, the consumer can choose another provider and start a new negotiation round. The used protocol in~\cite{RajavelT16} is an extension of the FIPA-CNP\footnote{see http://www.fipa.org/specs/fipa00029/}.
In~\cite{CoutinhoCSGJ16}, the authors introduce their own protocol by defining three states of messages - a concrete sequence which defines the message exchange according to this protocol is not described. The authors of~\cite{rajavel2015optimizing} use the Agent Communication Language as a protocol. It is maintained by the FIPA\footnote{http://www.fipa.org/repository/aclspecs.html} and was updated in 2002. The number of negotiation rounds is fixed. The authors distinguish between accept, reject, and counteroffers.

The survey revealed that a couple of different protocols exist. This could be an indicator that existing protocols are either insufficient for bilateral multi-round negotiations in the cloud domain or not well-known in the scientific community. Indeed, a clear motivation for the introduction of a novel protocol is missing in the surveyed papers.

\section{Utility Evaluation}
\label{sec:utility evaluation}

The main pillar of a negotiation strategy is the evaluation of offers. It is used for ranking offers but also for creating counteroffers.
In economics, two types of utility functions are used: ordinal utility functions and cardinal utility functions. Ordinal utility functions return ordinal values while cardinal utility functions return metric values so that the strength of preference can be expressed ~\cite{baruwal2015autoslam}. None of the analyzed papers, except~\cite{PittlMS16}, explicitly mentioned the type of utility function. However, within our analysis, we identified a dominant approach for calculating the utility of a cloud service: First, utility functions for service characteristics, such as price or availability, are defined so that a utility value can be calculated for each service characteristic. In a second step, these utility values are summed to a total utility value using a weighted sum function. So the used utility function for calculating the utility value of the service is a weighted sum whereby $i$ represents a negotiated service characteristic, $U_{i}$ represents the utility function of a service characteristic $i$, $n$ is the number of negotiable service characteristics and $w$ represents the weight.
\begin{equation}
	U_{Total}= \sum_{i=1}^{n} w_{i} \cdot U_{i}
\end{equation}

The authors of~\cite{SonKHKC16,RajavelT16,RajavelT162,rajavel2017adslanf,SonS15,HollowaySS15,
NajjarBP17,DastjerdiB15} introduce an additional constraint: $\sum_{i=1}^{n} w_{i}=1$. Hence, the weights a directly comparable to each other and the utility functions of the service characteristics are normalized between 0 and 1.

For the utility functions $U_{i}$ - which represent the utility value of a characteristics - the authors of~\cite{SonKHKC16,GhummanSL16,SonS15} define a most desirable value $max_{i}$  as well as an acceptable value $min_{i}$ which is also termed~\emph{reservation value} in several of the papers. The used utility function of these three papers is shown in the following equation:
\begin{equation}
	U_{i}= U_{min,i} + (1-U_{min,i}) \cdot \left| \frac{v_{i}-min_{i}}{max_{i}-min_{i}} \right|
\end{equation}
The right part of the equation is a typical normalization function which returns $1$ if the value of the service characteristic $v_{i}$ is equal to the maximum desired value $max_{i}$. The function returns $0$ in cases in which  $v_{i}$ is equal to the minimum acceptable value $min_{i}$.
The value of $U_{min,i}$ has to be determined by the user of the utility function and is used to~\emph{differentiate the situations not reaching an agreement and reaching an agreement at the} minimum proposal~\cite{SonKHKC16}. Indeed, this value can be considered as an offset. For instance, if $U_{min,i}=0.5$ then the minimum value of $U_{i}$ is $0.5$. The authors of~\cite{SonKHKC16} suggest using a value of around $0.01$. 
 Ghumman et al. introduce a slightly different approach: In~\cite{GhummanSL16}, the utility function, as shown in the following equation, is recommended. Here, $U_{min,i}$ is the utility value of the service characteristic $i$ while $\Delta U$ is defined as the difference between $U_{max,i}$ and $U_{min,i}$: $\Delta U=U_{max,i}-U_{min,i}$. So also~\cite{GhummanSL16} uses a linear utility function. The utility functions do not necessarily return values between $0$ and $1$ as $\Delta U$ can take on any values. Nevertheless, for calculating the total utility of a service, the authors suggest using the previously described weighted sum function.
\begin{equation}
	U_{i}=U_{min,i}+\Delta U \cdot \left( \frac{v_{i}-min_{i}}{max_{i}-min_{i}} \right)
\end{equation}
In the work of Son et al., concrete utility functions for cloud services are presented
~\cite{SonS15}. The authors considered the following characteristics:
\begin{inparaenum}[(i)]
	\item price,
	\item position of timeslot,
	\item performance,
	\item availability and
	\item reliability.
\end{inparaenum}
For each characteristic, a separate utility function is defined, and the utility values are combined to a total utility value by using weights for these utility values, whereby the sum of weights is 1. A difference to the previous utility functions is that the  total utility function is a discontinuous function, as the following example shows
\begin{equation}
	U_{Total}=
	 \begin{cases}
    0 & \exists U_{i}|U_{i}=0 \\
    \sum_{i=1}^{n} w_{i} \cdot U_{i}& otherwise \\

\end{cases}
\end{equation}
If one of the service characteristics is not in the acceptable range, the utility value is zero, and so the utility of the complete service is zero. Also, for the utility functions representing a service characteristic, a discontinuous function is used, as the following equation shows.
\begin{equation}
	U_{i}=
	 \begin{cases}
    	U_{min}+(1-U_{min}) \cdot \frac{max_{i}-v_{i}}{max_{i}-min_{i}} & min_{i} \leq v_{i} \leq max_{i} \\
   		0 & otherwise \\
	\end{cases}
\end{equation}

In other scientific works such as~\cite{BaranwalKRV18,RajavelT16,RajavelT162,rajavel2017adslanf,HollowaySS15,
DastjerdiB15,alsrheed2014intelligent,YaqubYWKLJ14} the previously described offset ($U_{min}$) is not foreseen: Here, the authors suggest using only the normalization functions as a utility function, which is shown in the following equation.
\begin{equation}
	U_{i}=\frac{v_{i}-min_{i}}{max_{i}-min_{i}}
\end{equation}

The authors of~\cite{HollowaySS15} mention the computational simplicity as a benefit of this approach, which~\emph{does not limit the applicability of} it. Further, the authors assume that three characteristics are negotiated: price, execution time, and availability; however, the approach is foreseen to be used for further characteristics. Similarly, the authors of~\cite{DastjerdiB15} suggest using autonomous negotiation for the parameters
\begin{inparaenum}[(i)]
	\item hard disk
	\item CPU
	\item RAM
	\item cost
	\item availability and
	\item deadline
\end{inparaenum}
while the approach introduced in~\cite{YaqubYWKLJ14} foresees a negotiation of non-functional issues:
\begin{inparaenum}[(i)]
	\item availability
	\item performance and
	\item backup.
\end{inparaenum}

 Baranwal et al. use such a utility function and suggest using only price for evaluating a service~\cite{BaranwalKRV18}. Hence, in the following equation, WP stands for the worst price (maximum for consumers, minimum for providers) and BP stands for the  best price (minimum for consumers, maximum for providers):
\begin{equation}
	U_{Total}= \frac{WP - p}{WP-BP}
\end{equation}
Chen et al., as well as Abedin et al., use the identical utility function~\cite{chen2014negotiation,AbedinCG14}.
In addition to that, the authors of~\cite{BaranwalKRV18} introduce an extended utility function: here, a further component is added which represents the opinion of the opponent, termed $of$ in the following:
\begin{equation}
	U_{Total}= U_{Price} \cdot w_{1} + of \cdot w_{2}
\end{equation}
$w_{1}$ and $w_{2}$ represent weights and the sum of them is $1$. 
Consumers calculate the $of$ value based on the reputation of providers (providers with a high reputation get a high $of$ value). For providers, the value $of$ is inter alia based on consumer loyalty. A high value of $of$ leads to a high total utility of $U_{Total}$. The authors of~\cite{alsrheed2014intelligent} also use a normalization function for evaluating the utility. However, the equation is slightly different from the previous approaches, as shown in the following:
\begin{equation}
	U_{i}=\frac{eval(v_{i})}{Max(eval(v_{i}))}
\end{equation}
$eval(v_{i})$ represents the utility value of a certain service characteristic and $Max(eval(v_{i}))$ represents the maximum utility value which can be achieved. These two values are put in relation to calculate $U_{i}$ whereby the maximum value is $1$.

The authors of~\cite{baruwal2015autoslam}  suggest using two different types of utility functions.
A point based utility functions defines for each possible service characteristic value  (x,y,z) a utility value (a,b,c)  as the following equations show.
\begin{equation}
	U_{i}=
	 \begin{cases}
    	a & i=x \\
   		b & i=y \\
   		c & i=z \\
	\end{cases}
\end{equation}
On the contrary, a so-called numeric utility function defines intervals as the following equation shows.
\begin{equation}
	U_{i}=
	 \begin{cases}
    	a & i \leq x \\
   		b & x < i \leq y \\
   		c & y < i \leq z  \\
	\end{cases}
\end{equation}
It is described that a utility function does not need to cover the complete service space. Indeed, the authors describe that such utility functions are usually too complex. Therefore, it is recommended to combine different types of utility functions to create a representative utility function. In addition, utility assertions can be used to complement utility functions. An example is depicted in figure~\ref{fig:combinedUtility}. Here, a minimum utility assertion ($\ge$0.6) is combined with a numerical utility function.

The authors of~\cite{PittlMS15} consider autonomous negotiation for the following resources:
\begin{inparaenum}[(i)]
	\item processing power
	\item storage
	\item RAM and
	\item price.
\end{inparaenum}
 Similar to other approaches, they assume that the utility representing the complete service is created by combining the single utility values with a weighted sum function. In contrast to the previously described approaches, which assume that consumers and providers know a minimum and maximum value for each service characteristic, the users of this utility function need only to know a minimum. This minimum should represent the lower bound of a service characteristic - if it can not be reached, the complete service has no value for the consumer. Such utility functions are not standardized between values between 0 and 1 - they do not have an upper bound, as the following exemplary utility function shows. $-\infty$ represents that the service characteristic - and consequently the complete service - has no utility for the user of the utility function.
\begin{equation}
	U(consumer_{storage})=
	 \begin{cases}
   		-\infty & storage < MIN_{storage} \\
    	sqrt(storage) & otherwise \\
	\end{cases}
\end{equation}
While most of the approaches assume a linear relationship between the utility value and service characteristic the authors of~\cite{PittlMS15,PittlHMS17wetice} assume a non-linear relationship by using sqrt or ln functions: The authors argue that in basic economics a saturation effect is assumed: the more a consumer has of a certain resource, the less additional utility is created by assigning an additional resources to the consumers. This effect is called diminishing marginal utility and can be reflected with the sqrt or ln functions. As the utility functions for the service characteristics have a completely different value space, the sum of weights, for calculating the total utility, does not need to be 1 - it is not bounded and should be used to make the utility values comparable.

The utility function which is introduced in~\cite{IshikawaF15}  looks like the following:
\begin{equation}
	U_{Total}=\sigma^{i}v_{i} \cdot c_{i}(v_{i}) \cdot w_{i}
\end{equation}
The sum of weights has not to be $1$ - indeed, the authors present an example with a negative weight. $c_{i}$ represents a constraint function: for each value of a service characteristic, it returns a value for increasing or decreasing the utility. Therefore, the authors exemplarily suggest using a discontinuous function\footnote{The parameter $\sigma^{i}$ is not described in the paper~\cite{IshikawaF15}}. 

\begin{figure}
 \begin{center}
    \includegraphics[width=0.65\linewidth]{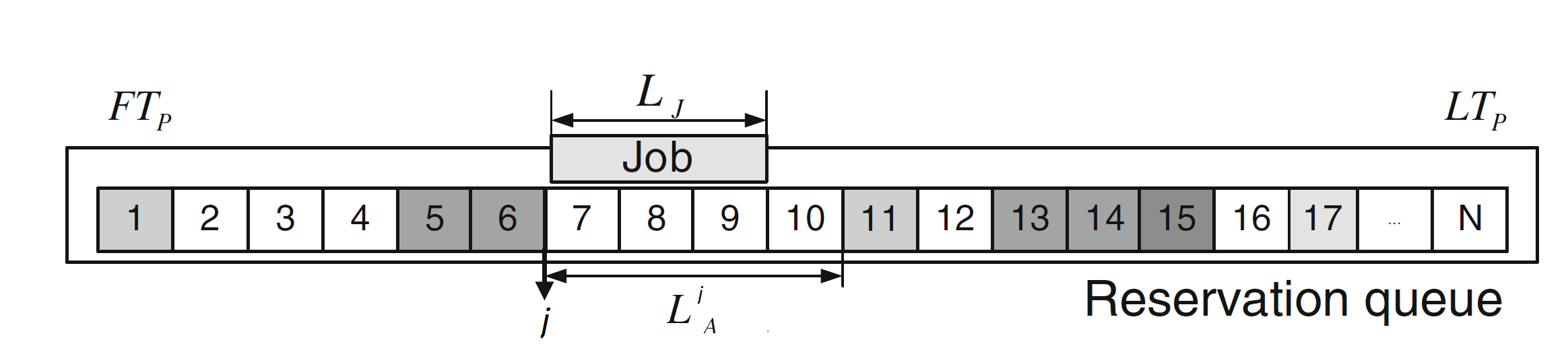}
    \caption{Timeslot mechanisms from\protect\cite{SonS15}}
     \label{fig:timeslot}
 \end{center}
\end{figure}

Son et al. further describe a concrete utility function for calculating the utility of the timeslot when a service is executed~\cite{SonS15}. It can be used by providers and is driven by two characteristics: early processing and fitting job size. These two characteristics are shown in the following equation. There are two weights: $w_{FF}$ is the weight for early job processing, while $w_{BF}$ is the weight for fitting job size. The first term is high for early tasks: $LT_{P}$ is the last timeslot considered by the provider, $FT_{P}$ is the first timeslot considered by the provider, while $j$ is the timeslot at which a task could be executed. Figure~\ref{fig:timeslot} visualizes these time points. The second term is high if a timeslot is available:   $L_{J}$ represents the requested job size while $L^{j}_{A}$ represents the length of a continuously available timeslot. For each time slot $j$, the corresponding value is calculated, and based on these values, a utility value is derived.
\begin{equation}
	U_{j}= w_{FF} \cdot \left( \frac{LT_{P}-j}{LT_{P}-FT_{P}} \right)   + w_{BF} \cdot \left( \frac{L_{J}}{L_{A}^{j}} \right)
\end{equation}

The authors introduced a timeslot utility function for consumers, which is generated based on given preferences. It is similar to the presented one, and so we do not describe it here. 

Also, other papers such as Rajavel et al., Ashok et al., or Najjara et al. use a utility function but do not describe them in detail~\cite{rajavel2015optimizing,NajjarBP17,AshokM15}. So, e.g., in~\cite{rajavel2015optimizing} the authors neither describe the sum of the weights nor the value range of the utility functions.

\begin{figure}
 \begin{center}
    \includegraphics[width=0.65\linewidth]{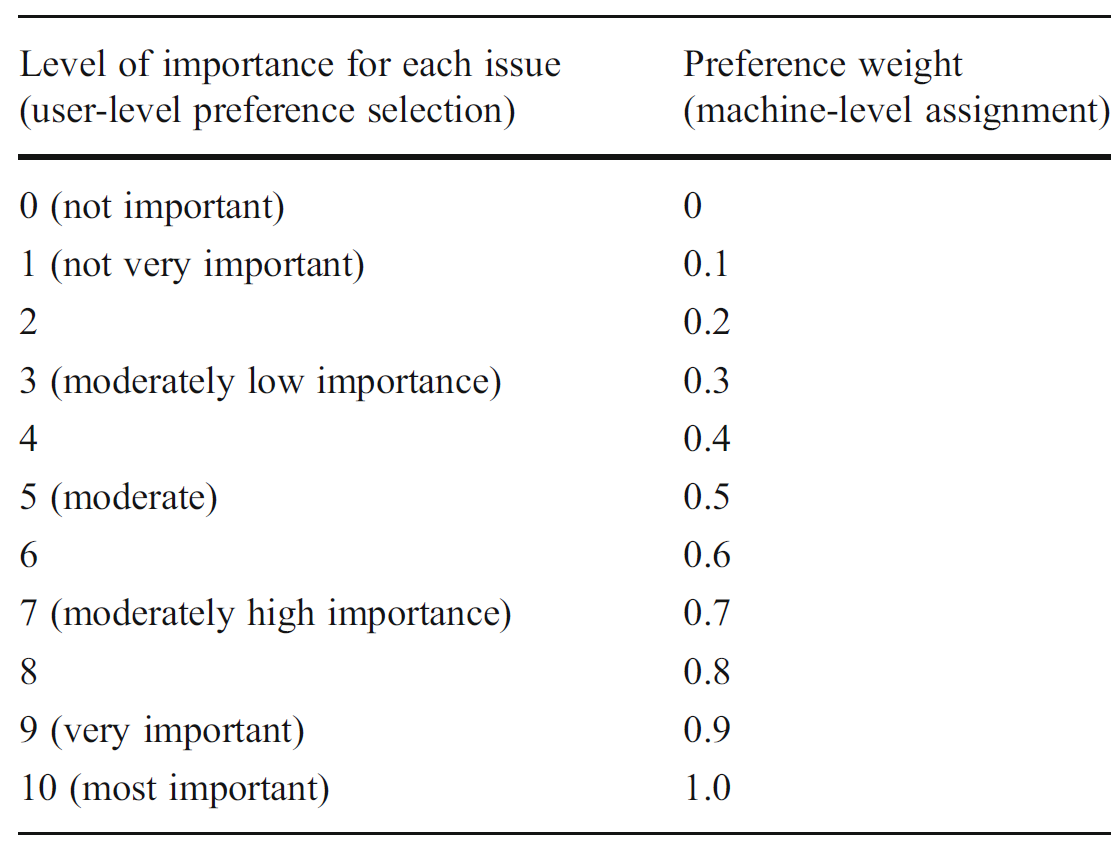}
    \caption{Interpretation of the weights used for calculating the utility of a service~\cite{SonS15}}
     \label{fig:parameterTable}
 \end{center}
\end{figure}

Independent of the used approach, a key challenge is to capture the appropriate weights for the utility functions. The authors of~\cite{SonS15} made a first step towards that direction and introduced a mapping of~\emph{user importance} and the weight parameters, as figure~\ref{fig:parameterTable} shows. No other paper describes an approach for setting the weights for the calculation of the utility of a service.

Table~\ref{tab:utilityEvaluationApproach} summarizes the findings: It is obvious that the normalization approach for calculating the utility of a service characteristic is dominant. In some works, the authors used a minimum utilization term $U_{min}$ while in other works, such a lower bound was not used. In both groups, a final utility value was created by summing up the utility values of the service characteristics. Few papers introduced other approaches or described the utility evaluation on an abstract level.

\begin{table}
\caption{Summary of utility evaluation approaches}
\centering
\begin{tabular}{p{5.2cm}p{7cm}}
\hline
\hline
 & Papers using the utility evaluation approach \\
\hline
\hline
Normalization with $U_{min}$ & \cite{SonKHKC16,SonS15,DastjerdiB15,GhummanSL16}  \\
Normalization without $U_{min}$ & \cite{BaranwalKRV18,RajavelT16,RajavelT162,rajavel2017adslanf,HollowaySS15,
alsrheed2014intelligent,YaqubYWKLJ14,chen2014negotiation,AbedinCG14,PittlMS16}  \\
Other approaches & \cite{PittlMS15,IshikawaF15,rajavel2015optimizing,baruwal2015autoslam,
AshokM15,PittlHMS17wetice}  \\
No approaches  & \cite{NajjarBP17,CoutinhoCSGJ16,dhanasekaran2015dynamic,GalI15,GhummanS17,
IlanyG16} \\
\hline
\end{tabular}

\label{tab:utilityEvaluationApproach}
\end{table}
\vspace{2em}

\section{Decision Making}
\label{sec:decisionMaking}

After offers are evaluated and ranked, a decision mechanism is necessary that defines how to respond to received offers. Indeed, this step of the negotiation process is described on the lowest level of detail. Most of the papers do not introduce concrete approaches for making decisions. For example, Chen et al. mention decision making in~\cite{chen2014negotiation} but do not provide details.

Rajavel et al. define two thresholds for making decisions~\cite{RajavelT162}: a lower threshold and a higher threshold. If the utility of an offer is lower than the lower threshold, then a reject message is created. If the utility of the offer exceeds the higher threshold, then an accept message is created. In all the other situations, a counteroffer using the behavioral learning system is created, as described in the following equation.
\begin{equation}
	Response=
	 \begin{cases}
    accept & \text{if } U_{Total} \geq T_{1} \\
    reject & \text{if } T_{2} < U_{Total}  \\
    counteroffer & \text{otherwise }  \\
\end{cases}
\end{equation}
The two thresholds $T_{1}$ and $T_{2}$ can be static values or dynamic values, which are, for instance, influenced by the utility values of received or sent offers. Indeed, the authors of~\cite{RajavelT162} make use of such dynamic thresholds. Also in~\cite{RajavelT16}, these two thresholds are used.  The introduced strategy in~\cite{PittlMS15} suggests using three thresholds in order to make decisions on received offers:
\begin{inparaenum}[(i)]
	\item accept active
	\item accept passive
	\item reject.
\end{inparaenum}
If the utility of an offer is lower than the~\emph{reject} utility threshold, then the offers are rejected. If a received offer in the acceptable state (a state of the WS-Agreement Negotiation protocol) has a utility which exceeds the~\emph{accept passive} threshold, then it is accepted. This is because in such a situation, the receiver can form an agreement immediately.
If an offer, which is not in the acceptable state, has a utility that exceeds the ~\emph{accept active} threshold, then the receiver responds with the offer in the acceptable state, which might lead to an agreement. There is no guarantee that an agreement will be formed. Usually, the accept active threshold is lower than the accept passive thresholds, but both thresholds can be identical.  


Dastjerdi et al. use the utility value of the received offers as a threshold, which the following equation shows~\cite{DastjerdiB15}.
\begin{equation}
	Response=
	 \begin{cases}
    	terminate & t_{offer} > t_{max} \\
   		\text{accept} & U_{offer} > U_{counteroffer} \\
   		\text{new counteroffer} & otherwise \\
	\end{cases}
\end{equation}

So an offer is only accepted if its utility is higher than the utility of the offer that would be used as a counteroffer. Dastjerdi et al. also explicitly consider the negotiation deadline in the decision-making process. Son et al. introduced a similar approach~\cite{SonKHKC16}: For the consumers as well as for providers, a negotiation deadline is assumed. For defining the threshold at which an offer is accepted, the authors use a time-dependent strategy considering a concession. Son et. al. describe concession as~\emph{the reduction in an agent's expectation based on its total utility}~\cite{SonS15}, which is in line with Ghumman et al., which summarizes concession as~\emph{moving towards its opponent's preferences}. In the following equation, $\tau$ represents the negotiation deadline and $\lambda$ represents the aggressiveness of the concession. $U_{total}^{t+1}$ is the threshold which has to be reached at time $t+1$ for an offer in order to be accepted:
\begin{equation}
	U_{total}^{t+1}= U_{total}^{t} - U_{total}^{t} \cdot \frac{t}{\tau}^{\lambda}
\end{equation}

Hence, the longer the negotiation takes, the lower the threshold. The utility of the received offers is ignored. Similarly, for making decisions, Najjar et al. suggest using a so-called aspiration rate ~\cite{NajjarBP17}. It is a threshold that represents the minimum utility that has to be reached by an offer. For this threshold, the authors suggest using a time-based concession as the following equation shows.

\begin{equation}
	AR=AR^{t-1} \cdot \left(\frac{t}{T}\right)^{\lambda}
\end{equation}
$\lambda$ is a parameter that determines the aggressiveness of the negotiation strategy,  $T$ represents the negotiation deadline, and $AR$ represents the aspiration rate. So the approach is almost identical to the approach introduced in~\cite{SonKHKC16}: Najjar et al. decrease the aspiration rate directly, while Son et al. increase the term that is subtracted from the current threshold to calculate the future threshold.
Najjar et al. recommend considering the concession of the negotiation partner for defining their own threshold. This would imply knowing the utility function of the negotiation partner. However, as the utility function of the negotiation partner is not known, the authors suggest using their own utility function for estimating the concession. Hence, the concession and so the threshold is calculated as follows, whereby $U_{i,t}$ and $U_{i,t-1}$ are utility values which are calculated with the own utility function:
\begin{equation}
	c=U_{i,t}-U_{i,t-1}
\end{equation}
So the threshold for accepting offers is currently reduced by $c$ or a value that is based on $c$. Additionally, it is suggested to use the opponent's concession not only to adapt the threshold but also to infer the negotiation deadline. A concrete approach was not the focus of~\cite{NajjarBP17}.

Other papers do not describe concrete rules for making decisions. So, e.g., in~\cite{BaranwalKRV18} it is described that if the utility of the received offer exceeds the expected utility, which is calculated using the utility function, then an agreement is created. If neither the deadline is reached nor the utility threshold is satisfied, then counteroffers are created.  For making decisions, the authors of~\cite{YaqubYWKLJ14} and~\cite{AbedinCG14} suggest using a threshold without describing details.
 For the decision making, the authors of~\cite{rajavel2015optimizing}  suggest using a Markov tree where the states represent negotiation states with the goal to optimize the reward. Thereby, a reinforcement approach is proposed - the reward is high if an agreement is formed, while it is low if a rejection is received.
The strategy introduced in~\cite{rajavel2017adslanf} proposes to use Behavior Learning for choosing the right behaviors based on negotiation history. The rewards function of the algorithm returns a high reward if an agreement is reached, and no reward if no agreement is reached. So the decisions of the strategy should lead to an agreement. It is described that the strategy should take an adequate action, whereby~\emph{action} is considered as a generic negotiation behavior. It could e.g. be a parameter for concession. The authors of~\cite{baruwal2015autoslam} suggest using classical rule-based approaches, which could be used for making decisions about how to respond to received offers.

\begin{table}
\caption{Summary of decision making approaches}
\centering
\begin{tabular}{p{4.2cm}p{8cm}}
\hline
\hline
 & Papers using the decision making approach \\
\hline
\hline
Thresholds & \cite{chen2014negotiation,RajavelT162,RajavelT16,PittlMS15,DastjerdiB15,SonKHKC16,
NajjarBP17,BaranwalKRV18,SonS15,YaqubYWKLJ14,AbedinCG14,PittlMS16}  \\
Other approaches & \cite{rajavel2015optimizing,rajavel2017adslanf,baruwal2015autoslam}  \\
No approaches & \cite{alsrheed2014intelligent,AshokM15,CoutinhoCSGJ16,dhanasekaran2015dynamic,
GalI15,GhummanS17,GhummanSL16,HollowaySS15,IlanyG16,IshikawaF15,
PittlHMS17wetice}  \\
\hline
\end{tabular}

\label{tab:decisionMakingApproaches}
\end{table}
\vspace{2em}

As table~\ref{tab:decisionMakingApproaches} shows, almost all identified approaches use thresholds to make decisions. These thresholds either consider the received offers, e.g., introduced in~\cite{DastjerdiB15}, or negotiation time, e.g., introduced in~\cite{SonKHKC16}. As mentioned in~\cite{PittlMS15}, other factors could also influence the thresholds, such as the number of parallel negotiations, which reflects the current market position: if, e.g., a consumer negotiates with a single provider, the threshold might be lower, as in situations where a consumer negotiates with multiple providers for the same service.

\section{Counteroffer Generation}
\label{sec:counteroffer}

The decision-making process might lead to the creation of counteroffers. The counteroffer generation process is summarized in this section.
During negotiation, the negotiation partners try to form an agreement. Therefore, the scientific literature uses the term~\emph{concession} too. In this context, concession is not used to reduce the threshold; it is used to increase the utility of generated offers for the negotiation partner. It is an important aspect that almost all the analyzed approaches foresee.

Abedin et al. suggest using a genetic algorithm for the creation of counteroffers~\cite{AbedinCG14}.  A detailed description of the genetic algorithm is not provided, but it is foreseen.
Also, the authors of~\cite{PittlMS16,PittlHMS17wetice} suggest using a genetic algorithm for counteroffer generation. Each individual of the genetic algorithm represents a potential counteroffer. The fitness function, which is used to select the fittest of a generation, is shown in the following.
\begin{equation}	
	\begin{aligned}
		F_{consumer}=U_{consumer}+\bar{U}_{provider} \cdot w
		\\
		F_{provider}=U_{provider}+\bar{U}_{consumer} \cdot w
	\end{aligned}
\end{equation}
The consumers use the value of their utility function ($U_{consumer}$) as well as an estimated utility function of the negotiation partner ($\bar{U}_{provider}$) for calculating a fitness value. Hence, the so-generated individuals with a high fitness value should have a high utility function for both the consumer and the provider. The estimated utility function of the negotiation partner is used because the real utility function is unknown. The weight $w$ is used to define the concession: if it is high, the utility function of the negotiation partner is important to achieve a high fitness value. If it is low, then the utility function of the negotiation partner is unimportant to achieve a high fitness value.

For creating counteroffers the authors of~\cite{PittlMS15} introduce a so called~\emph{naive} approach: Assuming that the negotiation partners do not know the preferences of each other, counteroffers are generated by randomly modifying e.g. two parameters of received offers: one parameter is modified so that the utility of the service for its sender is increased while the second parameter is modified so that the utility of the service for its sender is decreased. In total, the so-generated counteroffers should have a higher utility than the received offer. These counteroffers might have a higher utility for both consumers and providers as they have used utility functions.

 Son et al. use the negotiation deadline for concession~\cite{SonS15}. Therefore the authors suggest calculating a $\Delta U_{Total}$ which represents the concession at time $t$. The following equation is suggested:
 \begin{equation}
 	\Delta U_{total}=U^{t}_{total} \cdot \frac{t+1}{\tau}^{\lambda}
 \end{equation}

$\Delta U_{total}$ is then subtracted from the current utility target value of counteroffers which will be generated and used as counteroffers: $U^{t+1}_{total}=U^{t}_{total}-\Delta U_{total}$.  The counteroffer creation process is not described in detail in~\cite{BaranwalKRV18}  - it is stated that the price for the counteroffer is modified so that the target utility value is reached. According to this approach, it seems that only a single counteroffer is exchanged.

The negotiation process is structured by Hollowy et al. along two phases~\cite{HollowaySS15}:
\begin{inparaenum}[(i)]
	\item agreement phase and
	\item trade-off phase.
\end{inparaenum} In the agreement phase, the negotiation parties use a time-dependent strategy to exchange offers until one of the negotiation parties rejects the received offer (negotiation fails) or an initial agreement is found. Three concession strategies are mentioned, but are not described in the paper: linear, concede (great concession at the beginning), and boulware (great concession at the end). If an initial agreement is found, then the second phase starts, which focuses on modifying the offers so that their utility is higher for both the consumer and the provider. A concrete approach for counteroffer generation for the first phase is not given. For the second phase the offer is improved by using indifferent curves: an offer is modified by changing service characteristics based on the marginal rate of substitution (MRS)\footnote{the marginal rate of substitution is an exchange rate which expresses how many units of a certain service characteristic such as storage can be exchanged for a certain amount of other service characteristics such as RAM so that the utility of a market participant remains identical} which represents the slope of the indifference curve. The strategy suggests increasing a service characteristic $i$ as the following equation shows.
\begin{equation}
	v_{i}'=v_{i} \pm v_{i} \cdot \gamma_{i}
\end{equation}
$\gamma_{i}$ is the percentage of additional units with which service characteristic $i$ is improved. The modification of the service leads to a new marginal rate of substitution $MRS'_{i}$. The price for the newly generated service is defined as shown in the next equation.
\begin{equation}
	price'=price \pm \left( v_{i} \cdot \gamma_{i} \cdot MRS'_{i} \right)
\end{equation}
According to~\cite{HollowaySS15}, this can be rewritten to.
\begin{equation}
	MRS'_{i}=|MRS_{i}| \pm |MRS_{i}| \cdot p_{i}
\end{equation}
$p_{i}$ represents the size of the modification of the $MRS_{i}$ and might be changed over time. The algorithms try to interchange the service characteristics and price so that a Pareto-efficient outcome is reached.

The approach of Dastjeridi et al. for generating counteroffers is pure time-dependent as the following equation shows~\cite{DastjerdiB15}.
\begin{equation}
	O^{t}_{a \rightarrow b}[i]=
	 \begin{cases}
    	min_{i}+\alpha_{i}(t)(max_{i}-min_{i}) & \text{if $U_{i}$ is decreasing}  \\
    	min_{i}+(1-\alpha_{i}(t))(max_{i}-min_{i}) & \text{if $U_{i}$ is increasing}  \\
	\end{cases}
\end{equation}
$O^{t}_{a \rightarrow b}$ represents the generated counteroffer and $O^{t}_{a \rightarrow b}[i]$ represents the service characteristic $i$ of it. The first case is used if increasing $v_{i}$ increases $U_{i}$, while the second equation is used if increasing $v_{i}$ decreases $U_{i}$ (e.g., response time).
$\alpha$ is a variable that determines the concession. The authors of~\cite{DastjerdiB15} mention that polynomial and exponential functions could be used to calculate their value, as shown in the following equation. $\beta$ describes the tactic: if $\beta >1$, the creator has a high concession at the early stage of negotiation, while $\beta<1$ proposes an offer with its minimum requirements at the end of the negotiation. The value $k_{i}$ is the initial value with which the negotiation is started.
\begin{equation}
	\alpha_{i}(t)=
	 \begin{cases}
    	k_{i}+(1-k_{i}) \left( \frac{min(t,t_{max})}{t_{max}} \right)^{1/\beta} & Polynomial  \\
    	e^{1-min(t,t_{max})/t_{max}ln k_{i}} & Exponential  \\
	\end{cases}
\end{equation}
 For providers, the authors suggest using a modified counteroffer generation approach, which considers inter alia the utilization of the datacenter. The providers re-propose the price for the service characteristics described in the received offer - the other service characteristics of the counteroffer remain identical to the ones described in the received offer. The approach tries to increment prices of highly utilized resources while less utilized resources are sold at lower prices to consumers. The prices for the resources are calculated with the identical described equation; however, the $\alpha$ is modified according to the current utilization: if it is more utilized than the average resources, it is increased, otherwise it is decreased. To achieve such a behavior, $\beta$ is calculated dynamically:
 \begin{equation}
 	\beta_{i}=C_{1} \cdot e^{C_{2}(A_{i}-\bar{A})}
 \end{equation}
$C_{1}$ and $C_{2}$ are constants. $\bar{A}$ is the average utilization of all resources while $A_{i}$ is the current utilization of the resource $i$. Hence, $\beta$ is bigger if the resource is more utilized than the others and smaller if the resource is lower utilized than the others.

Also, the authors of~\cite{YaqubYWKLJ14} foresee a concession behavior during negotiation which depends on time and as the following equation shows.
\begin{equation}
	p=\frac{u-2ut_{c}+2(t_{c}-1+\sqrt{(t_{c}-1)^2+u(2t_{c}-1)}}{2t_{c}-1}
\end{equation}
$u$ represents the utility of the opponents offer, $p$ is the concession rate and $t_{c}$ is the current time. The value of $p$ increases for offers with low utilities. The concrete generation of the offers using the concession values is not described in the paper.

Son et al. describe that during a negotiation round, multiple offers are exchanged, which is called burst mode~\cite{SonS15}. The concrete number of counteroffers depends on the difference between the expected utility and the utility of the best offer that was received. If the difference is big, a low number of counteroffers is generated; if not, then a high number of counteroffers is generated. This is because the authors argue that additional offers do not lead to successful agreements if the utility difference is large. A concrete generation approach is not given in the paper. However, it is described that a similarity algorithm is used for selecting an appropriate counteroffer, as the higher the similarity is between the received offer and the counteroffer, the higher is the chance of forming an agreement. This is also stated in~\cite{PittlMS15}. 

Chen et al. consider three influencing factors for generating counteroffers, whereby the authors assume that only the price is negotiated~\cite{chen2014negotiation}:
\begin{inparaenum}[(i)]
	\item market competition
	\item time and
	\item opponent behavior.
\end{inparaenum}
For each influencing factor, a price for the counteroffer is determined, which is finally combined to a final price. The determination of the three prices is described in the following.
  Market pressure depends on the number of competitors $m_{t}$ and the number of opponents $n_{t}$. The authors of~\cite{chen2014negotiation} summarize these two numbers to a score as shown in the following equation:
\begin{equation}
	C(m_{t},n_{t})=1- \left( \frac{m_{t}-1}{m_{t}} \right) ^{n_{t}}
\end{equation}
This score is low if the number of competitors is high and the number of opponents is low. Based on the market pressure, the authors suggest generating the price for the next round, only the price is negotiated, as the following equation shows.

\begin{equation}
	P_{t+1}=P_{t} + (1-C(m_{t},n_{t})) \cdot (price_{max}-price_{min})
\end{equation}

So the price is updated based on the market competition.

The time, the second influencing factor, is also used for the concession as the following equation shows. Similar to the other approaches, $\lambda$ represents the aggressiveness of the concession while $T$ is the negotiation deadline. 
\begin{equation}
	P_{t+1}=P_{t} + \frac{1-[(t+1)/T]^{\lambda}}{1-(t/T)^{\lambda}} \cdot (price_{max}-P_{t})
\end{equation}

 Additionally, the opponent's history - the third influence factor - is used for determining the price. Here, the authors introduce different approaches - please see~\cite{chen2014negotiation} for more information. All the prices ($P_{1}, P_{2}, P_{3}$) proposed by the three influencing factors as well as the current price are combined to a final price by weighting the prices whereby the authors enforce that $\sum_{i}^n w_{i}=1$.
\begin{equation}
	P_{t+1}=P_{t} + w_{1} \cdot P_{1} + w_{2} \cdot P_{2} + w_{3} \cdot P_{3}
\end{equation}

The authors of~\cite{AshokM15} suggest using a linear function for the concession, which can be used for consumers and providers as the following equations show.

\begin{equation}
	concession_{\%}=1-\frac{min(\text{consumer utility},\text{provider utility})}{max(\text{consumer utility},\text{provider utility})}
\end{equation}

Here, the utility values of the negotiation partners have to be known. This is different from all the other approaches, which assume incomplete information. Further, the utility values have to be comparable.
The so calculated concession factor is used for determining the price of the counteroffer:

\begin{equation}
	price=price-(price \cdot concession_{\%})
\end{equation}


Similarly, the authors of~\cite{RajavelT162} describe the generation of counteroffers as the following equation shows. Here $f_{i}$ is the variable that defines the concession over time.
\begin{equation}
	v_{i}=min_{i}+(f_{i} \cdot (max_{i} - min_{i}))
\end{equation}
The variable  $f_{i}$  is defined as follows:
\begin{equation}
	f_{i}=C^{a} + \bigg((1-C^{a}) \cdot \Big(\frac{min(t,T_{1})}{T_{1}}\Big)^{1/\pi}\bigg)
\end{equation}
The authors distinguish between linear ($\pi =1$), conciliatory ($\pi =1/n$) and conservative concessions ($\pi =
1-\alpha$, $\alpha$ represents the maximum finite values of the negotiation attributes used in the offer. $C^a$ represents the first generated offer.

 Ghumman et al. introduce a negotiation strategy called~\emph{flip-flop}~\cite{GhummanSL16}.
For creating counteroffers, the authors suggest using a time-dependent utility function such as shown in the following equation:
\begin{equation}
	U_{i}=u_{min}+\Delta u \cdot \left( \frac{u_{i}-min_{i}}{max_{i}-min{i}} \right) -(\lambda \cdot t)
\end{equation}
The equation is identical to the previously explained utility function. However, an additional term is added to the end where $t$ represents the elapsed time and $\lambda$ represents a so-called~\emph {depreciation} factor: the authors suggest using e.g. 0.1 for it, and it increases the longer the negotiation takes. Hence, it is identical to the concession effect such as described in other literature. Also, the other values might change during negotiation. 
The authors try to predict the concession of the negotiation partner in order to boost the negotiation success. They suggest using the concession rate of the last three counteroffers to predict future concessions.  A concrete approach on how to create offers is not explained. Based on the user's behaviors, their own concession rate is adapted. Increasing is called a flip, while a flop represents a reduction of the concession rate. The overall goal is to predict the opponent's final offer so that it can be reached earlier in the negotiation process.

The authors of~\cite{baruwal2015autoslam} mention a rule-based approach but do not offer a concrete rule set. In~\cite{GhummanS17}, the authors describe an approach that sets the concession rate based on the estimated concession of the negotiation partner.

Creation of multiple offers at the same time is not considered in the strategy described in~\cite{SonKHKC16}. The work defines two termination conditions: either the deadline is reached or an agreement is formed. For the adaptation of counteroffers, the authors suggest an adaptive algorithm that is based on current workload trends. E.g., if a resource is rare, the weight is increased so that requests that do not require this resource are decreased. The concrete algorithm is based on a linear regression parameter, which is not described in the analyzed paper.

Table~\ref{tab:influenceFactors} summarizes the most important influence factors for creating counteroffers. Only one paper suggests using a pure random approach. All the other papers consider one or more influence factors for generating offers. Indeed, negotiation time is the most important influencing factor for the creation of counteroffers.

\begin{table}[htbp]
\caption{Summary of influence factors}
\centering
\begin{tabular}{p{4.2cm}p{8cm}}
\hline
\hline
 & Influence factors for counteroffer generation\\
\hline
\hline
Pure Random & \cite{PittlMS15}  \\
Time & \cite{PittlMS16,SonS15,DastjerdiB15,YaqubYWKLJ14,chen2014negotiation,RajavelT162,
GhummanSL16,SonKHKC16,PittlHMS17wetice}  \\
Received Offer & \cite{PittlMS16,HollowaySS15,SonS15,chen2014negotiation,GhummanSL16}  \\
Undefined/Other Factors & \cite{AbedinCG14,BaranwalKRV18,SonS15,chen2014negotiation,AshokM15,RajavelT162,
DastjerdiB15,GhummanSL16,SonKHKC16,GhummanS17} \\
No approach/Unknown & \cite{alsrheed2014intelligent,baruwal2015autoslam,CoutinhoCSGJ16,
dhanasekaran2015dynamic,GalI15,IlanyG16,IshikawaF15,NajjarBP17,RajavelT16,
rajavel2015optimizing,rajavel2017adslanf} \\
\hline
\end{tabular}

\label{tab:influenceFactors}
\end{table}
\vspace{2em}

Table~\ref{tab:counteroffers} summarizes how many counteroffers are exchanged during a negotiation round. A couple of papers such as~\cite{RajavelT16} try to reduce the number of offers which are exchanged while other approaches such as~\cite{PittlMS16} do not limit the number of exchanged offers.

\begin{table}[htbp]
\caption{Number of offers exchanged }
\centering
\begin{tabular}{p{4.2cm}p{8cm}}
\hline
\hline
& Offers exchanged during a negotiation round \\
\hline
\hline
\cite{AbedinCG14} & Single  \\
\cite{BaranwalKRV18} &  Multiple\\
\cite{chen2014negotiation} & Single\\
\cite{RajavelT162} & Single\\
\cite{RajavelT162} & Single \\
\cite{SonS15} & Multiple \\
\cite{SonKHKC16} & Multiple \\
\cite{HollowaySS15} & Single\\
\cite{DastjerdiB15} & Single\\
\cite{PittlMS15} & Multiple\\
\cite{PittlMS16} & Multiple \\
\cite{PittlHMS17wetice} & Multiple \\
\cite{alsrheed2014intelligent} & Single \\
\cite{GhummanS17} & Multiple \\
\cite{IshikawaF15} & Single \\
\cite{AshokM15} & Single \\
\cite{YaqubYWKLJ14} & Multiple \\
\hline
\end{tabular}

\label{tab:counteroffers}
\end{table}

\section{Scope and Formalization of Specifications}
\label{sec:scope}

Table~\ref{tab:scope} presents a summary of the scope of the negotiation strategies. Therefore, the steps of the aforementioned negotiation process are used. The table shows that the papers introducing negotiation strategies usually focus on parts of the negotiation process. Indeed, strategies that consider the complete negotiation process are rare. A couple of papers, such as~\cite{SonKHKC16,RajavelT16,RajavelT162}, additionally foresee a discovery process: providers register themselves in a central repository where consumers can query for them in order to start the negotiation.
Only in~\cite{PittlHMS17wetice} negotiation strategies for intermediaries are described - all other papers assume consumer-provider negotiations.

\begin{table}[htbp]
\caption{Scope of the introduced negotiation strategies}
\centering
\begin{tabular}{p{1.cm}llll}
\hline
\hline
Paper & Protocol & Utility Evaluation & Decision Making & Counteroffer Gen.   \\
\hline
\hline
\protect\cite{SonKHKC16} & \xmark & \cmark & \cmark & \cmark  \\
\cite{IlanyG16} & \cmark & \xmark & \xmark & \xmark  \\
\cite{GalI15} & \cmark & \xmark & \xmark & \xmark  \\
\cite{AbedinCG14} & \cmark & \cmark & \cmark & \cmark  \\
\cite{BaranwalKRV18} & \cmark & \cmark & \cmark & \cmark  \\
\cite{chen2014negotiation} & \cmark & \cmark & \cmark & \cmark  \\
\cite{RajavelT16} & \cmark & \cmark & \cmark & \xmark  \\
\cite{dhanasekaran2015dynamic} & \cmark & \xmark & \xmark & \xmark  \\
\cite{RajavelT162} & \cmark & \cmark & \cmark & \cmark  \\
\cite{SonS15} & \xmark & \cmark & \cmark & \cmark  \\
\cite{HollowaySS15} & \cmark & \cmark & \xmark & \cmark  \\
\cite{DastjerdiB15} & \cmark & \cmark & \cmark & \cmark  \\
\cite{baruwal2015autoslam} & \cmark & \cmark & \cmark & \xmark  \\
\cite{PittlMS15} & \xmark & \cmark & \cmark & \cmark  \\
\cite{PittlMS16} & \xmark & \cmark & \cmark & \cmark  \\
\cite{PittlHMS17wetice} & \xmark & \cmark & \xmark & \cmark  \\
\cite{alsrheed2014intelligent} & \xmark & \cmark & \xmark & \xmark  \\
\cite{GhummanSL16} & \cmark & \cmark & \xmark & \cmark  \\
\cite{IshikawaF15} & \cmark & \cmark & \xmark & \xmark  \\
\cite{CoutinhoCSGJ16} & \cmark & \xmark & \xmark & \xmark  \\
\cite{YaqubYWKLJ14} & \xmark & \cmark & \cmark & \cmark  \\
\cite{NajjarBP17} & \xmark & \xmark & \cmark & \xmark  \\
\cite{AshokM15} & \cmark & \cmark & \xmark & \cmark  \\
\cite{GhummanS17} & \cmark & \xmark & \xmark & \cmark  \\
\cite{rajavel2015optimizing} & \cmark & \cmark & \cmark & \xmark  \\
\cite{rajavel2017adslanf} & \cmark & \cmark & \cmark & \xmark  \\
\hline
\end{tabular}

\label{tab:scope}
\vspace{2em}
\end{table}

In~\cite{SonKHKC16}, it is described that providers face a trade-off between service qualities and prices: they aim at selling as many services as possible at high prices to consumers. At the same time, they try to avoid service violation penalties.  The proposed strategy encompasses an offer generator, a protocol, and a utility function. Further, the approach introduces a negotiation strategy manager as well as a proposal evaluator in addition to a discovery component. This setup is envisioned for both consumers and providers, whereby the authors assume that the service characteristics can be systematically prioritized for providers, but not for consumers, as their preferences reflect individual business needs. Providers have additional components, such as resource monitors as well as resource managers.  The strategy manager determines a concession rate while the proposal manager makes the decision regarding the acceptability of an offer.

The authors of~\cite{IlanyG16} conclude in their work that there is no optimal negotiation strategy for all situations. Therefore, participants should select the appropriate strategy from a set of predefined negotiation strategies. The selection approach is based on multiple technologies, such as logistic regression and linear regression. The approach introduced by the authors was evaluated using a test bed. The strategy is described on a high level of detail in the paper and assumes a concrete deadline as well as three offer states (proposal, accept, reject). For the evaluation, utility values are used, which are not described. 
The strategy focuses on providers only.

The negotiation strategy introduced in~\cite{AbedinCG14} is termed~\emph{agenda-based negotiation strategy} as the characteristics of a cloud service are negotiated sequentially. 
Further, it is assumed that the preferences of consumers are individual and therefore different for each consumer. The introduced strategy encompasses multiple steps:
\begin{itemize}
	\item Opinion collection: all consumers express the importance of the service characteristic using fuzzy linguistic terms.
	\item In a second step, fuzzy linguistic terms are normalized.
	\item Based on the previous two steps, group preferences are generated.
\end{itemize}
The negotiation itself is described as a process where offers and counteroffers are exchanged. If an agreement is found, then the next service characteristic is negotiated. If not, then a set of counteroffers is generated based on a genetic algorithm. 
In contrast to other approaches, the strategy foresees a systematic storage of offers as they are~\emph{possibly acceptable in the future}. 
 In the presented example, an exemplary utility function for evaluating offers is described, but no generic one. Further, thresholds are introduced for decision-making. 

Baranwal et al. ~\cite{BaranwalKRV18} assume that cloud consumers have a negotiation deadline until an agreement is necessary, but also providers have a deadline. Hence, a concession is suggested for both consumer and provider strategies.  The protocol the approach describes is underpinned by the assumption that consumers can respond with~\emph{a limited number} of offers. Both the consumer and the provider only negotiate the price, and both define a maximum price as well as a minimum price. 

Chen et al. consider a three-layer market~\cite{chen2014negotiation}: consumers, brokers, and suppliers. Consumers and providers do not negotiate directly; instead, the broker mediates negotiations between consumer and provider - these brokers are called~\emph{third-party brokers} in the paper. They are responsible for selecting requests and forming recommendations based on these requests.
The authors of the approach focus on reducing negotiation time and communication overhead. 
So the introduced strategy is based on two main pillars:  Classified Similarity Matching and Truncated Negotiation Group Gale Shapley Stable Matching. The first pillar focuses on avoiding similar or redundant negotiations, whereby the authors distinguish between situations where negotiation partners have complete, partial, or incomplete information. The Gale Shapley Stable Matching approaches were used for optimized matchmaking between consumers and providers.
A concrete counteroffer approach is missing, as well as a concrete utility function for cloud service characteristics.   The approach also foresees service discovery by introducing a UDDI. 

In~\cite{dhanasekaran2015dynamic}, a summary of negotiations in the cloud environment is presented as well as a prototype. However, the paper neither describes any details about a concrete negotiation strategy. Also, the prototype is presented on a very abstract level of detail.

In~\cite{RajavelT162}, the authors assume that negotiation partners do not know the preferences of the other negotiation partners. Therefore, they developed a negotiation strategy based on adaptive probabilistic behavioral learning. The approach is implemented using JADE, and so the agent communication language is used. However, a concrete protocol is not used. A UDDI is envisioned for service discovery. The offers exchanged during negotiation also consider the time slot when the expected service should be used. For making decisions on accepting an offer, the behavioral learning approach is described, which relies on rules of a probabilistic rule base. The used protocol distinguishes between offers, accepts, and rejects. Also, a deadline is determined. In a negotiation round, only one offer is exchanged.

The approach introduced in~\cite{SonS15} focuses on the negotiation of price, timeslot, and QoS parameters of cloud services. The authors argue that no negotiation strategy considers all three characteristics at the same time. The strategy foresees a negotiation deadline.

The authors of~\cite{HollowaySS15} assume a one-to-many negotiation setup for multiple services. Both the consumer and the provider express their requirements on a service using intervals that contain an upper bound as well as a lower bound for each service characteristic. The initial offer contains the values of the bounds as during the negotiation a concession mechanism is used.
 The authors further describe that Pareto-efficient offers are desirable. This can be reached with a trade-off mechanism: by improving a service characteristic in return for a reduction of another service characteristic.

The work presented in~\cite{rajavel2017adslanf} is an extension of~\cite{RajavelT15} and~\cite{RajavelT162}. In addition to total negotiation time and communication overhead, the strategy aims at optimizing each negotiation round by introducing a Reinforcement Learning method. Similar to the previous works, the JADE framework was used as a framework, and the messages conform to the ACL. The protocol encompasses accepts, rejects, as well as offers. 

\begin{figure}
    \centering
    \begin{subfigure}[b]{0.45\textwidth}
        \centering
    \includegraphics[width=0.9\linewidth]{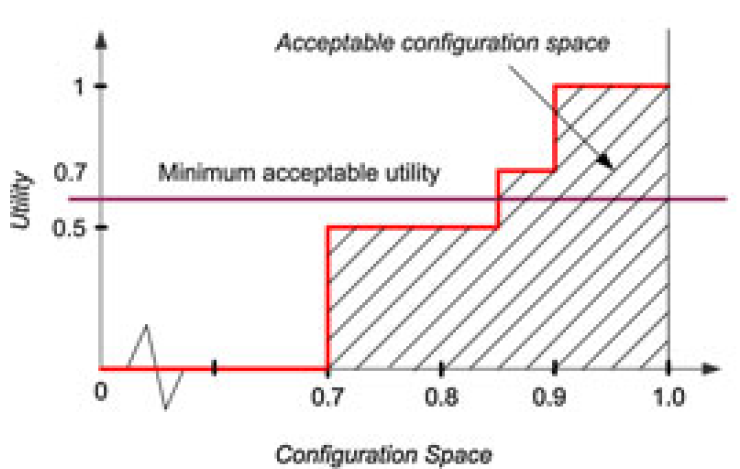}
    \caption{Example of a combined utility function introduced in~\cite{baruwal2015autoslam}}
     \label{fig:combinedUtility}
    \end{subfigure}
    \hfill
    \begin{subfigure}[b]{0.45\textwidth}
        \centering
        \includegraphics[width=0.99\linewidth]{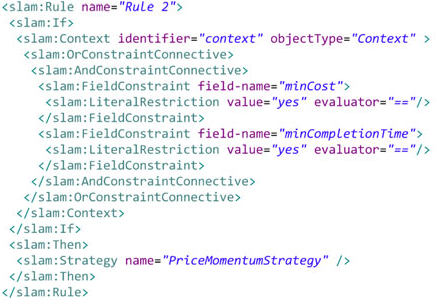}
        \caption{Examplary rule introduced in~\cite{baruwal2015autoslam}}
        \label{fig:exampleRules}
    \end{subfigure}
    \caption{Bilateral negotiation approaches}
\end{figure}

In~\cite{baruwal2015autoslam}, the authors present a generic negotiation framework called AutoSLAM. Even if the paper focuses on the description of the framework, several aspects of bilateral negotiation strategies were described, such as utility functions.
The paper also suggests a rule-based approach for defining negotiation strategies. With a focus on the AutoSLAM framework, a concrete strategy is not described. However, the authors suggest using classical~\emph{IF-Then} rules, which are well-known in the classical artificial intelligence domain. An example of such rules for selecting the appropriate negotiation strategy is depicted in figure~\ref{fig:exampleRules}. The introduced framework was not designed for a specific protocol - it can be used for multiple protocols.

The work in~\cite{PittlMS16} references~\cite{PittlMS15} regarding the utility functions and the decision rules. Both papers introduce a utility function as well as a counteroffer generation approach. Further, both papers are based on the WS-Agreement Negotiation protocol.

The work presented in~\cite{PittlHMS17wetice} introduces negotiation strategies for intermediaries. It is assumed that in future cloud markets, several participants will sell and purchase cloud resources. Intermediaries are market participants that neither run datacenters, such as providers, nor use the services, such as consumers. Hence, the author's strategies for resellers are categorized into three classes:
\begin{inparaenum}[(i)]
	\item neutral strategy
	\item buy-first-strategy
	\item sell-first-strategy.
\end{inparaenum}

According to the first strategy, consumers and providers form agreements at the same time. The intermediary has almost no risk as it simply forwards the offers of the provider and the consumer. According to the buy-first-strategy, the intermediary purchases services at temporarily low prices and tries to sell them later. On the contrary, according to the sell-first-strategy, the intermediaries first make an agreement with consumers and in a second step, they try to find appropriate providers. For the neutral strategy, different markup strategies can be used whereby the markup is the profit of the intermediary~\footnote{please see ~\cite{PittlHMS17wetice} for more information about markup}
\begin{inparaenum}[(i)]
	\item According to the fixed-markup strategy, the intermediary tries to achieve a fixed markup, such as e.g. $3\$$.
	\item According to the proportional-markup strategy the intermediary tries to achieve a fixed share of the price as a markup such as e.g. $3\%$.
	\item According to the time-dependent-markup strategy, the intermediary uses a time-dependent strategy for the markup. So, e.g., the intermediary starts the negotiation by planning with a high markup. During negotiation, it reduces the markup.
\end{inparaenum}

Possible negotiation strategies are described on a high level of detail in~\cite{alsrheed2014intelligent}. For instance, for a strategy called~\emph{HardHeaded} it is described that~\emph{the target of the learning module is to learn the utility value and the weights of the opposing agent}. Details on counteroffer generation or on how to make decisions are not described.


The authors of~\cite{IshikawaF15} mention that a cloud provider faces high volatilities. Therefore, negotiation strategies based on a static utility function are not appropriate. Hence, the authors advise strategies that consider the current utilization of their data centers. The paper also describes that concessions should be used for the negotiation. However, a concrete approach is missing.

In~\cite{YaqubYWKLJ14}, a time-restricted negotiation strategy is introduced. Thereby, the authors follow a rule-based approach. The implementation of the paper is based on the negotiation framework GENIUS. 

A strategy for providers is introduced in~\cite{NajjarBP17}. The authors emphasize two key characteristics of it:
\begin{inparaenum}[(i)]
	\item Adaptive: the negotiation behavior changes during negotiation.
	\item Open: Market participants can enter and leave the negotiation anytime.
\end{inparaenum}
The negotiation strategy was implemented using the framework EMan.

The authors of~\cite{CoutinhoCSGJ16} envision not only bilateral but also multilateral negotiations. The work focuses on the description of negotiation based on Interaction Abstract Machines. Thereby, it neither introduces concrete decision rules, utility functions or concession strategies.


Ashok et al. suggest using a weighted sum function for calculating the total utility~\cite{AshokM15}. Details about the weight or the calculation of individual utility values are not described.

Ghumman et al.~\cite{GhummanS17} introduce a SLA Management approach based on the negotiation strategy presented in~\cite{GhummanS16}.

The authors of~\cite{rajavel2015optimizing} inter alia consider service discovery as a significant part of their framework. Thereby, providers register themselves in a repository so that consumers can find them.

Table~\ref{tab:implementation} summarizes the frameworks and software environments that are used in the scientific community for implementing the introduced negotiation strategies. It shows that JADE, CloudSim, as well as customized implementations are widely used for implementing the strategies.

\begin{table}[htbp]
\centering
\begin{threeparttable}
\caption{Summary of frameworks used for implementing the strategy}
\begin{tabular}{lp{9.5cm}}
\hline
\hline
& Papers using this Platform \\
\hline
\hline
JADE & \cite{SonKHKC16,RajavelT16,RajavelT162,rajavel2017adslanf,SonS15}  \\
CloudSim & \cite{chen2014negotiation,DastjerdiB15,PittlMS15,PittlMS16,PittlHMS17wetice} \\
Repast Simphony\tnote{1} & \cite{NajjarBP17} \\
Customized & \cite{IlanyG16,BaranwalKRV18,dhanasekaran2015dynamic,MaciasG16,alsrheed2014intelligent,
GhummanSL16,GhummanS17,CoutinhoCSGJ16,AshokM15,YaqubYWKLJ14,YaqubYWKLJ14}  \\
\hline
\end{tabular}
\begin{tablenotes}
\item[1] http://repast.sourceforge.net
\end{tablenotes}
\vspace{2em}
\label{tab:implementation}
\end{threeparttable}
\end{table}


Table~\ref{tab:formalization} lists the formalization techniques that the scientific community uses for describing negotiation strategies. It is obvious that the textual description of them (informal) is dominating. One approach (\cite{dhanasekaran2015dynamic}) completely describes the negotiation strategy with text. All the other specifications make use of at least one further formalism. Also, formulas, e.g., for describing utility functions and counteroffer creation approaches, are widely used. Some of the papers also use diagrams, such as flow charts, for specifying the negotiation process. UML diagrams seem not to be widely used for the description of negotiation strategies - only one paper makes use of a sequence diagram.
 Algorithms, which describe negotiation strategies at least in pseudo-code, are also used by some papers.

\begin{table}[htbp]
\caption{Formalization Technique}
\centering
\begin{tabular}{p{3cm}p{9cm}}
\hline
\hline
& Used Formalism \\
\hline
\hline
\cite{SonKHKC16} &  UML Sequence Diagram, Formulars, Informal   \\
\cite{IlanyG16} &  Informal, Formular   \\
\cite{GalI15} & Informal, Formular   \\
\cite{AbedinCG14} &  Informal, Flow-Charts \\
\cite{BaranwalKRV18}  & Informal, Formulars, Algorithm \\
\cite{chen2014negotiation} & Informal, Flow-Chart, Formulars \\
\cite{RajavelT16} & Algorithm, Informal, Formulars \\
\cite{dhanasekaran2015dynamic} & Informal \\
\cite{RajavelT162} & Informal, Algorithm \\
\cite{rajavel2017adslanf} & Informal, Formulars \\
\cite{SonS15} & Informal, Formulars \\
\cite{HollowaySS15} & Informal, Formulars, State-Diagram \\
\cite{DastjerdiB15} & Informal, Sequence Diagram, Formulars \\
\cite{baruwal2015autoslam} & Informal, Formulars,Flow-Charts \\
\cite{PittlMS15} & Informal, Formulars, Algorithm\\
\cite{PittlMS16} & Informal, Formulars \\
\cite{PittlHMS17wetice} & Informal, Formulars \\
\cite{alsrheed2014intelligent} & Informal, Formulars \\
\cite{GhummanSL16} & Informal, Formulars, Flow-Charts, Aglorithm \\
\cite{IshikawaF15} & Informal, Formulars \\
\cite{CoutinhoCSGJ16} & Informal, Formulars \\
\cite{AshokM15} & Informal, Formulars, Algorithm \\
\cite{YaqubYWKLJ14} & Informal, Formulars, Algorithm \\
\hline
\end{tabular}

\label{tab:formalization}
\end{table}

\section{Quality Assurance and Potential Weaknesses}
\label{sec:quality}

During the survey process, a couple of threats occurred, which we summarized in table~\ref{tab:surveyCriteria}. In the analysis step, the exclusion of relevant publications due to an inappropriate search method is a significant threat. So we executed a couple of countermeasures:
\begin{inparaenum}[(i)]
	\item We documented the corpus in section~\ref{sec:corpa} and~\ref{sec:corpb} so that the reader can track the excluded papers.
	\item During our work, we discussed the process and our findings with other research peers on our faculty as well as at summits.
	\item Finally, we tried to avoid a bias in our survey by considering generic keywords (\emph{bilateral negotiation strategy}).
\end{inparaenum}

The main challenge of analysis B was the correct identification of~\emph{relevant} publications. We considered all papers that describe bilateral multi-round negotiation strategies for cloud SLAs. As already described in a previous section, some papers foresee a negotiation approach but do not introduce it. Such papers were not included in the survey; borderline papers (class III) were discussed.

Finally, during the survey, a main threat is the misinterpretation of concepts that are described in scientific publications. Unclear concepts were discussed to reduce this threat.


\begin{sidewaystable}
\caption{Summary of the analysis steps}
\footnotesize
\centering
\begin{tabular}{llll}
\hline
\hline
& Analysis A & Analysis B & Analysis C \\
\hline
\hline
Selection Criteria & \textbf{Exclusion Criteria:} & \textbf{Inclusion Criteria:} & \textbf{not relevant} \\
&  Published before 2014  &  relevant for the survey  &  \\
&  No scientific publication (e.g. patent or book)  &  (bilateral multi-round negotiation for cloud-SLA)  & \\
&  Not relevant  & & \\
Review Method &  Reading the title  &  Reading the title &  Reading the paper\\
 &  Cross-Reading the abstract  &  Reading the abstract &  Peer-discussion of the paper\\
 &  Reading the keywords  &  Reading the keywords & \\
 &  Reading the conference/journal title  &  Reading the conference/journal title & \\
  &    &  Cross-Reading the paper & \\
Executor & 2 persons (PhD student,PostDoc)  & 2 persons (PhD student, PostDoc) & 3 persons (PhD student, \\
 &  & &  PostDoc, Professor)\\
\hline
Threats &  exclusion of relevant publications &  subjective inclusion criteria &  misinterpretation of concepts  \\
 &  inadequate search method & & \\
Quality Assurance Measures &  discussion with peers  &  discussion with peers  &  discussion with peers\\
 &  clear documentation of corpus  &  check if~\emph{cloud-SLA} is explicitly mentioned  & \\
  &  usage of generic keywords  & & \\
\hline
\end{tabular}

\label{tab:surveyCriteria}
\end{sidewaystable}

\section{Recommendations}
\label{sec:discussion}

The survey focused on the analysis of autonomous bilateral negotiation strategies. During the survey, we identified the following issues, which represent a baseline for future research work.
\begin{itemize}
	\item  \textbf{Heterogeneity of Protocols.}

A widely accepted standard protocol or de facto standard protocol for autonomous multi-round bilateral negotiations does not exist in the scientific community. Hence, negotiation strategies have a different scope and are not directly comparable. For instance, a decision mechanism that considers three different states of offers is hard to compare to another decision mechanism that considers, e.g., 8 different states of offers. 
	
	\item  \textbf{Simplistic Utility Functions.}
	
Most of the introduced papers use a weighted sum function as a utility function. Thereby, utility values for each characteristic are calculated, which are summarized to a single utility value using a weighted sum function. None of the surveyed approaches presents a mechanism on how to capture the weights used in the weighted sum function, which is a hindrance to their adoption in industry. Advanced approaches such as the Prospect Theory - which are considered to be more appropriate to represent human perceptions~\cite{Farokhi14} - are not considered yet for bilateral negotiations.
	
	\item  \textbf{Different Scope.}	
	
Negotiation strategies are complex mechanisms, so most researchers focus on specific aspects of a negotiation strategy, e.g., on the counteroffer generation. Descriptions of complete negotiation strategies, which would be necessary to implement them in industry, are rare. Hence, there is a need for the definition of interfaces so that the different components introduced in the scientific community could be bundled to form a complete negotiation strategy.
	
	\item  \textbf{Degree of Formalization.}		

Our survey shows that informal text is heavily used for describing negotiation strategies. However, for implementing scientific algorithms in industry, more formal techniques are necessary. This would also foster comparability between the introduced approaches.	

	\item  \textbf{Lack of Implementations.}		

The identified negotiation strategies are only prototypically implemented. In almost none of the papers, the implementation is publicly available and maintained as an open-source project. Hence, it is often unclear how the strategy behaves as scientific papers focus on the description of the most important aspects.

	\item  \textbf{Missing KPIs.}		

KPIs along which negotiation strategies could be compared and evaluated are necessary in order to boost bilateral multi-round negotiations. The need for KPIs for comparing strategies is underpinned by the assumption of~\cite{IlanyG16}, who conclude in their work that there is no single optimal negotiation strategy for all situations.

\item  \textbf{Heterogeneity of Visions.}	

	 The survey at hand shows that different visions exist regarding the provision of negotiation strategies. Some work assumes that the negotiation component is provided by third parties, while other works envision that each participant implements its own strategy. Further, some papers assume consumer-provider markets while others foresee intermediaries. Hence, the goals of the negotiation strategies are diverging, and so a comparison between them is hard to achieve. Creators of negotiation strategies have to define for which market structure they optimized their strategy.
	
\end{itemize}

%

\section{Conclusion}
\label{sec:conclusion}

Bilateral multi-round negotiations are a promising approach for trading Cloud services in future Cloud markets. In the paper at hand, we surveyed negotiation strategies that the scientific community introduced. Thereby, we found out that weighted-sum functions for the evaluation of offers are widely accepted in the scientific community. However, approaches for decision making are rare: most of the papers use threshold values and do not foresee optimization techniques for improving decision making. For the generation of counteroffers, the scientific community foresees different methods ranging from a pure random generation to approaches that use genetic algorithms. During the generation, influence factors such as time, received counteroffers, or market pressure are considered. The factor~\emph{time} has the highest acceptance in the scientific community. The scope of the introduced negotiation strategies is broad: while some publications try to describe complete strategies, most of the papers focus on a few aspects. For the description, the scientific community heavily uses informal natural language, which makes the comparability between the approaches difficult.

Based on the survey, we derived a couple of recommendations. Currently, one of the biggest issues is the limited comparability of negotiation strategies due to the different scopes. Here, the scientific community has to establish a framework and interfaces so that the different components, which are introduced in scientific publications, can be composed so that a comprehensive negotiation strategy emerges. Further, widely accepted KPIs are necessary so that the performance of the negotiation strategies can be compared. Within our survey, we found out that the adoption of emerging  Blockchain and Smart Contract technologies in negotiation strategies is limited. This is a promising research direction in order to improve, inter alia, the integrity of the negotiation process.


\bibliographystyle{splncs03}
\bibliography{allCites, addon}

\appendix

\section{Corpus A}
\label{sec:corpa}

\begin{itemize}

\item Papers found using the search term \emph{Bilateral Negotiation Strategies}:
\newline
\cite{IlanyG16} \cite{BaarslagHHJ16} \cite{HaoL14} \cite{ChenW14} \cite{CaoLLD15} \cite{leu2015development} \cite{ZhangRZ15} \cite{ZhengMBX14} \cite{vij2015negotiation} \cite{HaberlandML17} \cite{LinKBTHJ14} \cite{BaarslagHHDJ14} \cite{ShyurS15} \cite{SongH00L016} \cite{leu2015optimize} \cite{RajavelT16} \cite{DastjerdiB15} \cite{Fujita14} \cite{huffmeier2014being} \cite{RenZ14} \cite{ZafariM14} \cite{GalI15} \cite{YaqubYWKLJ14} \cite{aydougan2017alternating} \cite{MansourK14} \cite{ChenW15}  \cite{rajavel2015optimizing} \cite{ChenHWZZ15} \cite{KolomvatsosTH15} \cite{HaniPH15} \cite{CoutinhoCSGJ16} \cite{RanaldoZ16} \cite{MellLG15} \cite{heese2015single} \cite{BalachandranM15} \cite{BaarslagHJ14} \cite{HsuKHL16} \cite{GratchNJ16} \cite{HashmiMNAM16} \cite{ChenHWTL14} \cite{AydoganBHJY15} \cite{patrikar2015approach} \cite{PittlMS16} \cite{AbedinCG14}

\item Papers found using the search term \emph{cloud Negotiation Strategies}:
\newline
\cite{DastjerdiB15} \cite{ZhengMBX14} \cite{YaqubYWKLJ14} \cite{RajavelT16} \cite{FalasiSH15} \cite{GhummanSL16} \cite{omezzine2015towards} \cite{SonS15} \cite{SonKHKC16} \cite{ShojaiemehrRQ18} \cite{BaranwalKRV18} \cite{HaniPH15} \cite{rajavel2015optimizing} \cite{RanaldoZ16} \cite{peng2017dynamic} \cite{Sim15} \cite{ChenACGLT15} \cite{IshikawaF15} \cite{chen2015cpn} \cite{HabesBV17} \cite{KerteszKB14} \cite{HollowaySS15} \cite{rajavel2017adslanf} \cite{AshokM15} \cite{WangCW16} \cite{PittlMS16} \cite{OmezzineSTC16} \cite{dastjerdi2015cloudpick} \cite{NajjarBP17} \cite{wang2015sla} \cite{GhummanS16} \cite{RajavelT15} \cite{DastjerdiB14} \cite{NajjarBP17a} \cite{WangWS16} \cite{mansour2015approach} \cite{PittlMS162} \cite{GhummanS17} \cite{awasthi2016multi} \cite{Ravindran15} \cite{chen2014negotiation} \cite{shyam2018resource} \cite{StavrinidesK17} \cite{sim2016agent} \cite{CoutinhoCSGJ16} \cite{HuangKM15} \cite{baruwal2015autoslam} \cite{dhanasekaran2015dynamic} \cite{MaciasG16} \cite{PittlMS15} \cite{HafezE16} \cite{VelosoMB15} \cite{ManoharanS15} \cite{wazir2016service} \cite{ScocaUN17} \cite{SinghC16} \cite{alsrheed2014intelligent} \cite{MansourK14} \cite{rajesh2015genetic} \cite{SyedKSKBKS17} \cite{RajavelT162}

\item Papers found using the search term \emph{Bazaar Negotiation Strategies}:
\newline
\cite{PittlMS16} \cite{Mach17} \cite{PittlMS15} \cite{PittlMS15} \cite{PittlMS16} \cite{PittlMS16EDOC} \cite{PittlHMS17wetice} \cite{ZhangRZ15} \cite{SyedKSKBKS17} \cite{ZhangRZ14} \cite{PittlMS16} \cite{LinKBTHJ14}

\end{itemize}

\section{Corpus B}
\label{sec:corpb}

\begin{itemize}

\item Papers which were identified as relevant for the survey:
\newline
\cite{ShojaiemehrRQ18} \cite{HafezE16} \cite{BaarslagHHJ16} \cite{SinghC16} \cite{IlanyG16} \cite{SonKHKC16} \cite{AbedinCG14} \cite{BaranwalKRV18} \cite{chen2014negotiation} \cite{RajavelT16} \cite{dhanasekaran2015dynamic} \cite{RajavelT162} \cite{rajavel2017adslanf} \cite{GalI15} \cite{SonS15} \cite{HollowaySS15} \cite{DastjerdiB15} \cite{baruwal2015autoslam} \cite{PittlMS15} \cite{PittlMS16} \cite{PittlHMS17wetice} \cite{alsrheed2014intelligent} \cite{GhummanSL16} \cite{GhummanS17} \cite{IshikawaF15} \cite{CoutinhoCSGJ16} \cite{AshokM15} \cite{YaqubYWKLJ14} \cite{rajavel2015optimizing} \cite{NajjarBP17}

\end{itemize}

\end{document}